\documentclass[preprint2,times,tighten]{aastex6}
\pdfoutput=1 

\usepackage{amsmath,amstext}
\usepackage[all]{hypcap} 
\usepackage{graphicx}
\usepackage{placeins}
\usepackage{float}
\usepackage{lipsum}

\usepackage[normalem]{ulem}
\usepackage{color}

\newcommand{\kgsinsert}[1]{\textcolor{black}{\textbf{#1}}}
\newcommand{\kgsdel}[1]{\textcolor{red}{}}
\newcommand{\kgsdelete}[1]{\textcolor{red}{}}

\newcommand{\jcbins}[1]{\textcolor{black}{#1}}
\newcommand{\jcbdel}[1]{\textcolor{pink}{}}

\newcommand{\kojins}[1]{\textcolor{black}{#1}}
\newcommand{\kojdel}[1]{\textcolor{red}{}}

\shorttitle{Spectroscopic Binary Fraction in IN-SYNC}
\shortauthors{Jaehnig et al.}

\newcommand{\eg}{{\rm e.g.}}

\newcommand{\ie}{{\rm i.e.}}                                   

\newcommand{\kms}{\ensuremath{\mathrm{km\:s}^{-1}}}
\newcommand{\ms}{\ensuremath{\mathrm{m \cdot s}^{-1}}}
\newcommand{\vsini}{\ensuremath{\mathrm{\it{v} \sin \it{i}}}}
\newcommand{\drv}{\ensuremath{\Delta\mathrm{RV}}}
\newcommand{\rvmin}{\ensuremath{v_\mathrm{min}}}
\newcommand{\rvmax}{\ensuremath{v_\mathrm{max}}}
\newcommand{\ndrv}{\ensuremath{\mathrm{NDRV}}}
\newcommand{\fndrv}{\ensuremath{NDRV}}
\newcommand{\teff}{\ensuremath{T_\mathrm{eff}}}
\newcommand{\logg}{\ensuremath{\log g}}
\newcommand{\LH}{\ensuremath{P (O | X_f, I)}}
\newcommand{\pdetecti}{\ensuremath{p_{\mathrm{detect},i}}}
\newcommand{\msun}{\ensuremath{M_{\odot}}}
\newcommand{\sigdrv}{\ensuremath{\sigma_{\Delta RV}}}
\newcommand{\sigrv}{\ensuremath{\sigma_{RV}}}

\begin{document}

\title{\textbf{IN-SYNC. VII. Evidence for a decreasing spectroscopic binary fraction \kgsinsert{from 1 to 100 Myr} within the IN-SYNC sample}}
\author{Karl Jaehnig\altaffilmark{1,2},
        Jonathan C.\ Bird\altaffilmark{2,6},
        Keivan G.\ Stassun\altaffilmark{2,1},
        Nicola Da Rio\altaffilmark{3},
        Jonathan C.\ Tan\altaffilmark{3,4},
        Michiel Cotaar\altaffilmark{5},
        and Garrett\ Somers\altaffilmark{2,7}
        }
\email{karl.o.jaehnig@vanderbilt.edu}
\begin{center}
\altaffiltext{1}{Department of Physics, Fisk University, Nashville, TN, 37208, USA}
\altaffiltext{2}{Department of Physics and Astronomy, Vanderbilt University, Nashville, TN 37235, USA}
\altaffiltext{3}{Department of Astronomy, University of Florida, Gainesville, FL 32611, USA}
\altaffiltext{4}{Department of Physics, University of Florida, Gainesville, FL 32611, USA}
\altaffiltext{5}{Department of Clinical
Neurosciences, University of Oxford, Oxford, United Kingdom}
\altaffiltext{6}{Stevenson Fellow}
\altaffiltext{7}{VIDA Fellow}
\end{center}
\begin{abstract}
We study the occurrence of spectroscopic~binaries in young star-forming regions using the INfrared Spectroscopy of Young Nebulous Clusters~(IN-SYNC) survey, carried out in SDSS~III with the APOGEE spectrograph. Multi-epoch observations of thousands of low-mass stars in Orion~A, NGC~2264, NGC~1333, IC~348, and the Pleiades have been carried out, yielding H-band spectra with a nominal resolution of R=22,500 for sources with H\textless12~mag. Radial~velocity precisions of $\sim$0.3 \kms~were achieved, which we use to identify radial velocity variations indicative of undetected companions. We use Monte~Carlo simulations to assess the types of spectroscopic binaries to which we are sensitive, finding sensitivity to binaries with orbital periods $\lesssim 10^{3.5}$~d, for stars with $2500~{\rm K} \le \teff \le 6000~{\rm K}$ and \vsini \textless 100~\kms. Using Bayesian inference, we find evidence for a decline in the spectroscopic binary fraction, by a factor of 3--4 from the age of our pre--main-sequence sample to the Pleiades age . The significance of this decline is weakened if spot-induced radial-velocity jitter is strong in the sample, and is only marginally significant when comparing any one of the pre--main-sequence clusters against the Pleiades. However, the same decline in both sense and magnitude is found for each of the five pre--main-sequence clusters, and the decline reaches statistical significance of greater than 95\% confidence when considering the pre--main-sequence clusters jointly. Our results suggest that dynamical processes disrupt the widest spectroscopic binaries ($P_{\rm orb} \approx 10^3 - 10^4$~d) as clusters age, indicating that this occurs early in the stars' evolution, while they still reside within their nascent clusters.
\end{abstract}

\keywords{Keywords to be added later } 
\maketitle

\section{Introduction} \label{sec:intro}
It is known that most stars form in clusters \citep{ladalada2003} and that a substantial fraction of solar-type stars in the field are binaries \citep[e.g.,][]{DM1991,raghavan2010}. Thus it is expected that most solar-type stars must form in binaries \citep[e.g.,][]{lada2006} and that the binary fraction in star-forming clusters should be at least as high as that observed in the field.
\par 
At the same time, the binary fraction is not expected to be a static quantity. Wide binaries in particular are not expected to last long in dense stellar regions due to the larger number of interactions between systems, as well as the smaller binding energy of the binary. Indeed, \citet{ghez1993} used speckle imaging to suggest that the occurrence of solar-type binaries with separations of hundreds of AU declines by a factor of $\sim$3.5 from the pre--main-sequence (PMS) to the main sequence. Tight binaries ($a < 20~AU$) are expected to be able to last longer due to their higher binding energy. \cite{mason1998} studied the binary fraction among Gyr-aged populations and found, using chromospheric activity, that the binary fraction was decreasing with age on Gyr timescales. 
To properly understand the binary fraction and therefore how stars form in populations, we must understand where this dynamical processing of the binary fraction begins to take place, for both wide and tighter (spectroscopic) binary systems. 
Estimates for the spectroscopic binary fraction at field ages are typically $\sim$10\% \citep[e.g.,][]{DM1991}.
It is necessary to also characterize the spectroscopic binary fraction of star forming regions to establish the possible evolution of the spectroscopic binary fraction during the PMS phase. 
Finally, the spectroscopic binary fraction especially at young ages is important for understanding planet formation in binary systems, including circumbinary planets around tight binaries and the dynamical evolution of planets around individual stars in wide binaries \citep[see, e.g.,][]{kraus2015, kraus2016}.

\par
Observing such young stellar systems at ages of a few Myr can be challenging, since not all of these systems will have cleared the gas and dust around them. This gas and dust obscures the stars in the optical. For example, in the Orion Nebula Cluster the visual extinction, $A_V$, can range from a few tenths of a magnitude to tens of magnitudes, depending on the sightline and distance into the cloud \citep[see, e.g.,][]{hillenbrand2007, dario2014}.Infrared observations are able to penetrate through the gas and dust so that spectroscopic measurements of radial velocity variations may be performed to identify spectroscopic binaries \cite[see, e.g.,][]{prato2002a, prato2002b}.
\par
The Sloan 2.5m telescope \citep{gunn2006} feeds light to the Apache Point Observatory Galactic Evolution Experiment(APOGEE) spectrograph \citep{majewski2017}. The APOGEE spectrograph primarily observes in the H-band (1.51$\mu$m-1.7$\mu$m) of the near-infrared spectrum. The spectrograph can observe up to 300 source targets per plate, with a 2 arcsecond fiber size, and a nominal spectral resolution of 22,500. The APOGEE survey looked at 100,000 stars within the Milky Way galaxy, focusing on red giants. The Infrared Spectroscopy of Young Nebulous Clouds (IN-SYNC) survey was an ancillary project during SDSS-III that used the APOGEE spectrograph to carry out  high volume, high precision observations of pre-main sequence stellar populations. The IN-SYNC survey provides a very powerful and unique opportunity to study binary systems and the binary fractions of young star forming populations as well as the kinematics of the youngest star forming regions in multiple types of star forming environments having carried out observations within the galactic bulge, halo, and disk.  

In this paper, we measure and compare the spectroscopic binary fraction of five PMS clusters and the MS Pleiades cluster using the multi-epoch IN-SYNC data. Binary candidates are identified via RV-variability. Straightforward comparison of the raw cluster binary fractions would be misleading due to significant cluster-to-cluster differences in observational cadence and target sampling. After careful accounting of these observational differences and uncertainties within a probabilistic model of the binary fraction within each cluster, we find that the spectroscopic binary fraction for the main sequence Pleiades cluster is a factor of $\sim$3--4 less than the average spectroscopic binary fraction for the PMS clusters. We attribute this to the dissolution of relatively wide (orbital periods 10$^2$--10$^4$~d) spectroscopic binaries, not well studied previously among PMS clusters, that are probed by our sample. 

We discuss the data we use in this paper in Section 2, go over the processes employed to characterize the candidates with an unseen companion in Section 3, discuss the Bayesian Inference framework we used to correct for observational differences in Section 4, discuss the derived cluster binary fraction distributions in Section 5, frame our results within a broader scientific context in Section 6, and state our conclusions in Section 7.

\section{Data and Sample}\label{sec:data}

\subsection{The IN-SYNC Survey and APOGEE} \label{ssec:data_overview}
\begin{figure*}
\epsscale{1.2}
\plotone{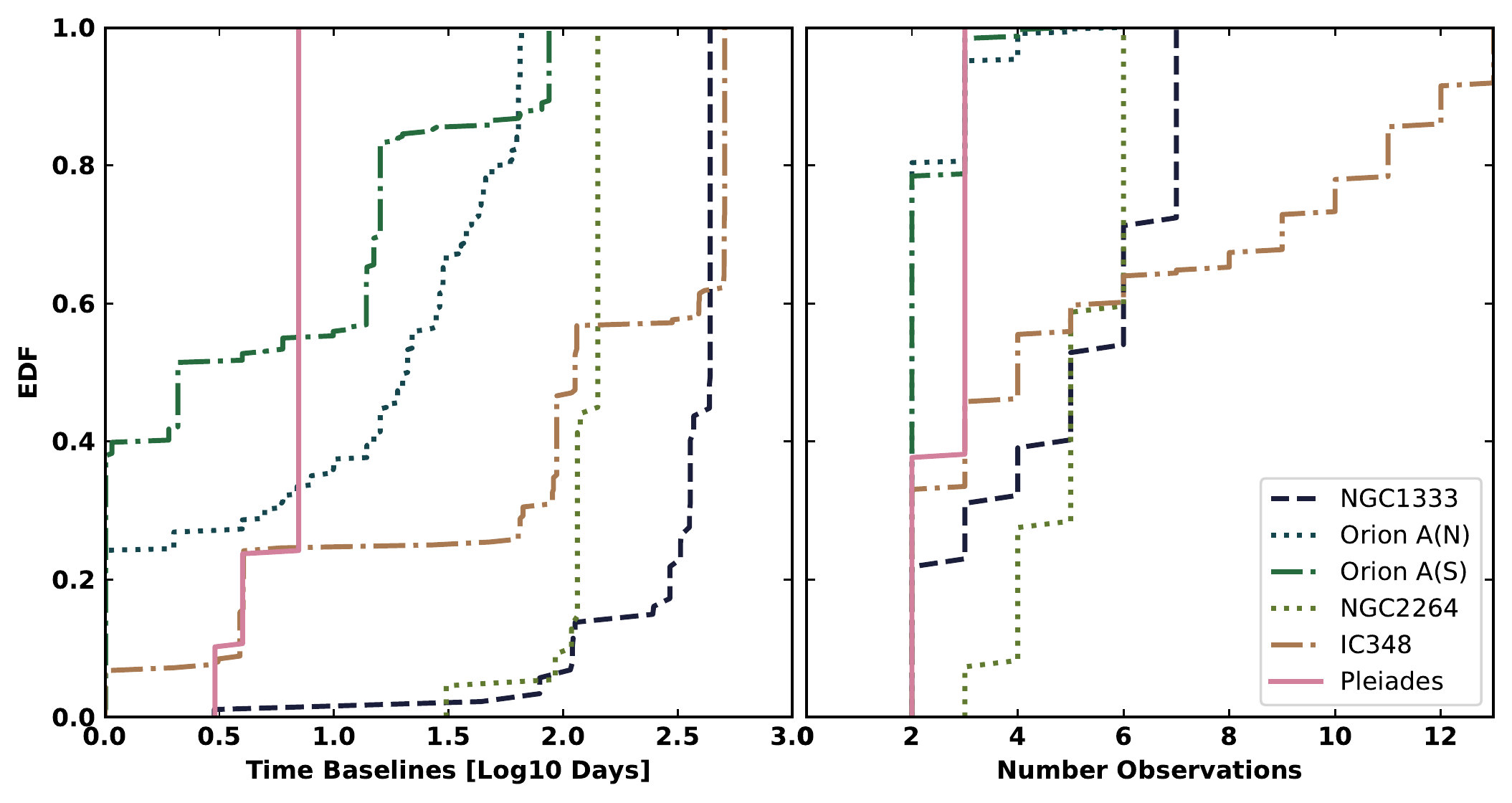}
\parbox{18cm}{\caption{ Left Panel: Empirical distribution functions of the observational time baselines in $log_{10}$ days for the 5 IN-SYNC clusters and the Pleiades. Right Panel: Empirical distribution functions of the number of epochs per observation for the 5 IN-SYNC clusters and the Pleiades. The values within both these panels are for the final vetted sample of stars from Section \ref{ssec:stellar_systematics} }} 
\label{time_baseline_nepochs}
\end{figure*}

The IN-SYNC survey was an ancillary project to the SDSS III APOGEE program \citep{IN-SYNC_paper_1}. Using the APOGEE spectrograph, the IN-SYNC survey obtained multi-epoch high-resolution (R=22500) spectroscopy of dust obscured young star forming regions. Multiple regions were observed during the IN-SYNC survey, consisting of IC348, and NGC1333 in the Perseus Cloud, NGC2264, and the Orion A molecular cloud complex (see Figure~\ref{time_baseline_nepochs}). The IN-SYNC team has already performed numerous analyses into these regions. \cite{IN-SYNC_paper_2} looked at the kinematics of the embedded pre-main sequence population of NGC1333. \cite{IN-SYNC_paper_3} looked at the dynamical state of IC348. \cite{IN-SYNC_paper_4} looked the Orion A Molecular Cloud Complex and compared derived stellar parameters to previous literature catalogs, and most recently \cite{IN-SYNC_paper_5} looked at the kinematics and dynamical state of the Orion A population. 
\par 
The IN-SYNC survey obtained multi-epoch spectra for over 3000 pre-main sequence stars. The IN-SYNC survey derived stellar parameters for the Pleiades which were observed with the APOGEE spectrograph \citep{ahn2014}, and whose spectra were processed using the same IN-SYNC data reduction pipeline. The IN-SYNC data reduction pipeline uniformly derived  effective temperatures, surface gravities, radial velocities, rotational velocities, H-band veiling, and corresponding stellar parameter errors. For full details of the IN-SYNC pipeline reduction process, please see \cite{IN-SYNC_paper_1} (Section 3.1.1 -- 3.1.3).

\subsubsection{Cluster membership of study sample}
\label{ssec:membership}
The raw IN-SYNC pipeline catalog contains derived stellar parameters for 12945 individual observations of 4771 stars. Not all of these observed stars necessarily belonged to any of the five observed regions and so cross-matching of 2MASS ids were used between the observed targets and literature catalogs to separate true members of IC348, NGC1333, NGC2264, Orion A, and the Pleiades from observed field stars. \kojdel{We initially demand that every observation in our raw dataset have a signal to noise $\ge$ 20, thus removing 131 stars, with 1412 observations.}

For NGC1333 we used the \cite{rebull2015} catalog of known NGC1333 members and cross-matched their locations within the \cite{cutri2003} 2MASS point source catalog to extract member 2MASS IDs. \kojdel{Within this set of 854 ids we then found 88 stars that were science targets observed for IN-SYNC with derived stellar parameters for 291 observations.} The NGC1333 cluster has been estimated to have an age of $\sim$1 Myr \citep{gutermuth2008}, which we employ here.

\par 
For NGC2264 we used the final cross-matched set of 2MASS ids composed from the catalogs of \cite{lamm2005}, \cite{makidon2004}, \cite{sung2008}, and \cite{sung2009} by the IN-SYNC team. \kojdel{and found 114 stars to be members of NGC2264 with derived stellar parameters for 636 observations.} The age of the NGC2264 cluster has been cited to be around $\sim$3 Myr \citep{venuti2014}.

For the Orion A molecular cloud cross matching was performed using a provided catalog (Dr. Nicola Da Rio, Private Communication) of membership lists composed from Optical Spectra (\cite{hillenbrand1997}, \cite{fang2009}, \cite{fang2013}, \cite{hsu2012}, \cite{hsu2013}, \cite{dario2012}), Infrared excess (\cite{getman2005}, \cite{pillitteri2013}), and X-ray (\cite{megeath2012}).  \kojdel{This catalog provided us with 2691 stars, from which we found 3594 stars that were observed science targets for IN-SYNC with derived stellar parameters for 3580 observations.} It is known that the Orion A molecular cloud is home to numerous different stellar populations. From \cite{IN-SYNC_paper_4}, the argument is made that given the approximate size of the entire Orion A cloud of 40pc, as well as its relatively young age, it is not possible for dynamical interactions to homogenize the ages of the young stellar populations throughout the filament. We therefore cannot look at the cloud as one cluster and decide to separate it two smaller, sub-clusters.

\par
Using \logg \ -- \teff \ isochrones, \cite{IN-SYNC_paper_4} finds that there is distinct separation in the ages of the Orion A region around $\delta \sim 6^{\circ}$. \cite{IN-SYNC_paper_5} also finds this distinction between populations occurring at $\delta \sim -6^{\circ}$ when looking at the position-position-velocity space of the population. We therefore decide to split the Orion A region into two sub-clusters which we study separately.  We delineate the northern sub-cluster as all systems with a declination of $\delta \ge -6^{\circ}$. This region, which we call Orion A(N), is dominated by the Orion Nebula Cluster. \kojdel{, with 1360 stars and 2309 observations.} The southern sub-cluster is designated as all systems with a declination of $\delta < -6^{\circ}$, which we call Orion A(S). \kojdel{, with 719 stars and 1285 observations.} Following \cite{IN-SYNC_paper_4}, we use the  \logg \ -- \teff \  estimated ages of $\sim$1.5 Myr and $\sim$2.5 Myr for Orion A(N), and Orion A(S), respectively.

\par 
For IC348, we cross matched identified members within the catalogs of \cite{luhman1999}, \cite{luhman2003}, \cite{lada_et_al2006}, and \cite{muench2007} to develop a set of member 2MASS IDs. \kojdel{From this set we find 198 stars within IC348 that were observed science targets within IN-SYNC with derived stellar parameters for 808 observations.} \cite{bell2013b} recently found the IC348 region to be $\sim$6 Myr, which we use for our population sample.

\par
Finally within the Pleiades we used the online catalog of identified cluster members from \cite{IN-SYNC_paper_1}. \kojdel{to find 1425 Pleiades cluster members with 2MASS ids. From this set we find that there are 68 stars that were observed IN-SYNC science targets with derived stellar parameters for 202 observations.}
While the age of the Pleiades has been well understood to be $\sim$100~Myr\citep{meynet1993}, the exact age has been difficult to constrain. Using Lithium depletion and K-band photometry, \cite{martin2001} found the age of the Pleiades to be $\sim$115 Myr. \cite{stauffer1998} found the age of the region to be $\sim$125 Myr using Keck-II spectroscopic data. We decide to employ the median age of 115 Myr for this population sample from \cite{martin2001} 


\begin{figure*}
\epsscale{1.2}
\plotone{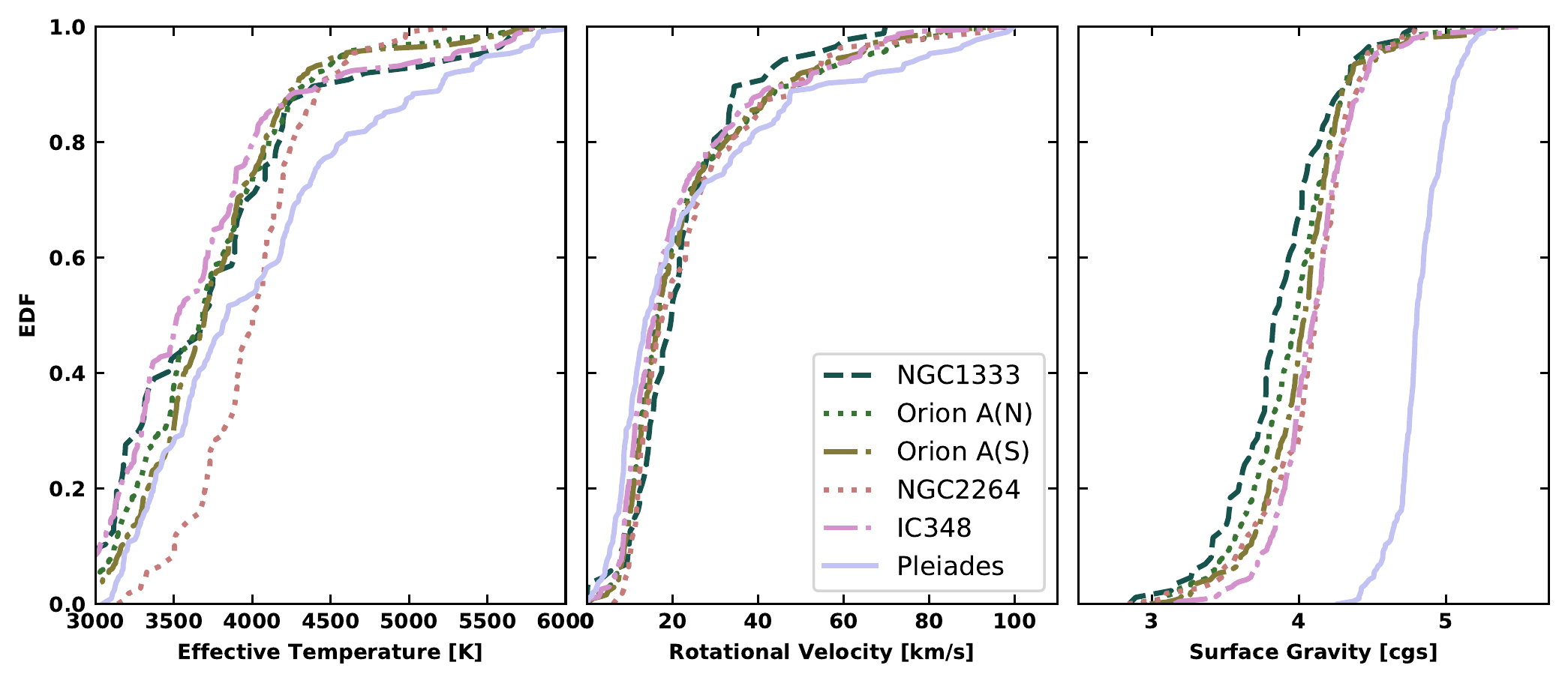}
\parbox{18cm}{\caption{Empirical distribution functions of \teff (left panel), \vsini (center panel), and \logg (right panel) for the final vetted sample of the five IN-SYNC clusters and the Pleiades. The difference in the EDFs of the surface gravity within the Pleiades compared to the 5 IN-SYNC clusters is due to the inherent age difference between the Pre-main sequence clusters and the Main-sequence clusters.}} 
\label{stellar_parameters_edfs}
\end{figure*}

\subsubsection{Data Quality Assurance}
\label{ssec:parameters}
Having identified all stars that are cluster members, we proceed to ensure that the derived stellar parameters are accurate. We are motivated to drop all individual observations with a signal to noise below 20 since these observations were too noisy to provide accurate information on stellar parameters (see \cite{IN-SYNC_paper_3}, Section 2.3). Stars with only one observation are also removed since they offer no information for looking at radial velocity variability.

Over 1600 stars have had repeat observations during the IN-SYNC survey. Spectral parameters of interest-- \logg, \teff, \vsini, RV, and SN -- were determined for each individual observation and reported as per epoch parameters within the IN-SYNC catalog. We identify candidate binary systems by measuring the variation in radial velocity associate with each star (see Section~\ref{sec:sb_search}). Following \citet{IN-SYNC_paper_1} and \citet{IN-SYNC_paper_4}, we looked at the per epoch variation of \vsini, \logg, \teff\ and found that highly variable per epoch \teff\ measurements for any single system were the most significant indicator of problems with the data extraction and thus of spurious, non-binary radial velocity variability within the IN-SYNC observations. Therefore, we looked to the per epoch measurements of effective temperature for each star and remove those epochs with \teff\ measurements that deviate significantly beyond what would be expected from standard gaussian error. 

In about 75\% of the stars observed within our data, we have fewer than 4 epochs of measurements. Therefore we calculate the Median Absolute Deviation (MAD) for each set of measurements. The MAD is a robust statistic, compared to the standard deviation. To use the MAD as an estimator for outlier rejection similar to the standard deviation, we employ a constant scale factor, which changes depending on the underlying distribution.

For a gaussian distribution, this scale factor is 1.4286, and 1.4286$\times$MAD can be used to designate the interquartile range of values assuming a gaussian distribution \citep[see][]{huber_ronchetti2009}. We calculate the MAD value for the set of \teff measurements from stars with more than 2 epochs of observation and remove epochs with parameter deviations exceeding 3 times the value of 1.4286$\times$MAD, which is equivalent to demanding that the stellar parameters lie within 3$\sigma$ of their distribution. 

For stars with only 2 epochs of observation we cannot rely on the MAD statistic, as both measurements of \teff\ are equidistant from the calculated median of the set and would not be removed as outliers using the MAD, even if the two values of \teff\ were highly variable. Instead we calculate the $\chi^{2}$ of the set of two measurements of \teff, and find the probability of that $\chi^{2}$ value within a $\chi^{2}$  distribution with 1 degree of freedom. We reject stars with a probability less than $10^{-3}$ as this implies both measurements came from separate distributions.

We also perform \teff\ and \vsini\ cuts to ensure that the radial velocity variability within the IN-SYNC sample is due to an unseen companion and not the result of poor spectral fitting. We drop stars whose median \teff\ and/or \vsini\ measurements do not fall in the range of $ 2500K \le \teff\ \le 6000K$, and $\vsini \le 100\kms$, respectively. We implemented these cuts following \cite{IN-SYNC_paper_3} and summarize their reasoning for these cuts as follows:
\begin{itemize}
  \item The IN-SYNC survey did not go deep enough to observe cool stars ($<2500K$), and such low \teff\ fits indicate noisy fits within the pipeline.
  \item IN-SYNC stars with \teff\ $>6000K$ and \vsini\ $>100\kms$ have hydrogen lines in their spectra that affected radial velocity derivation, making them too noisy and unreliable.
\end{itemize}

From an initial raw sample of 12945 observations of 4771 stars, we removed non-cluster members using 2MASS ids, implemented a signal to noise cut on individual observations, ensured that sets of observations did not deviate from standard gaussian behavior within \teff\, and removed stars that were too cool/too hot and too rapidly rotating. 

Our final vetted catalog of data contains 4642 measurements for 1418 stars. Table~\ref{tab:bin_frac} contains a tally of the number of stars in each cluster included in the analysis that follows. The empirical distribution functions (EDFs), which gives the fraction of a sample for a variable that are at or below any value of the measured variable, of the time baselines, number of observations, and stellar parameters for the final measurements that result from the steps described above are summarized in Figures~\ref{time_baseline_nepochs} and \ref{stellar_parameters_edfs}.

\section{Identifying Binary Systems in IN-SYNC } \label{sec:sb_search}

The IN-SYNC survey achieved radial velocity precision down to 0.3~\kms, which in principle make these data well suited to study multiplicity within these young star forming regions.
Temporal evolution of a star's radial velocity is a clear indication of an orbiting companion \citep{iben_tutukov1996}. Most studies use RV variation to both identify stars with unseen companions and reconstruct the orbits of the binary system. Orbit-reconstruction requires several to many measurement epochs to ensure adequate phase coverage. With multiple epoch measurements, binary candidates can be identified by the RMS scatter of radial velocities, or any other variance measure \citep[\eg,][]{troup2016}. However since many of the stars observed within IN-SYNC are pre-main sequence stars, with high temperatures, and high rotational velocities, we expect their radial velocity RMS scatter to be primarily driven by \vsini\, which has been seen before amongst young F spectral type stars \citep[\eg,][]{galland2005}. Indeed checking the radial velocity RMS of the IN-SYNC stars against \vsini, we find a Kendall's Rank Correlation Coefficient of 0.315, with p-value $<<$ 0.05. We therefore cannot rely on the radial velocity RMS as an effective measure of binarity. 

\par 
\citet{fernandez2017} have recently presented orbit solutions for a small number of spectroscopic binaries in the APOGEE IN-SYNC data for which a relatively large number of epochs were available. We do utilize this sample to check that our metric for binary candidate identification (see below) is reliable. However, here we are not concerned with orbit-reconstruction, only binary candidate identification, and we also cannot rely on radial velocity scatter to categorize a system as a spectroscopic binary. We therefore employ a simple, robust measure of RV variability between any pair of epochs which we develop below. Given a set of $N\geq2$ RV measurements $\{v_n\}$ of a star, the maximum variation in radial velocity is \drv$=\rvmax\ - \rvmin$, where \rvmax$= max(\{v_n\})$ and \rvmin$= min(\{v_n\})$. Each radial velocity measurement $v$ has an associated measurement error $\sigma_v$. We can determine the statistical significance of \drv\ via comparison with its uncertainty $\sigma_{\drv}$, propagated from the individual measurement errors $\sigma_{\rvmax}$ and $\sigma_{\rvmin}$ of \rvmax\ and \rvmin, respectively. We thus define the Normalized Delta RV (\ndrv) as the RV variation, \drv, normalized by the error $\sigma_{\drv}$, 
\begin{equation}
NDRV = \frac{\drv}{\sigma_{\drv}} = \frac{ \rvmax - \rvmin}{\sqrt{\sigma_{\rvmax}^{2} + \sigma_{\rvmin}^{2}}}.
\label{eq:ndrv}
\end{equation}
\ndrv\ is  our primary statistic for identifying spectroscopic binary candidates. 

\subsection{Investigation of NDRV behavior with SN}
\label{ssec:ndrv_sn_correction}
\par
\begin{figure*}[!ht]
\begin{center}
  \includegraphics[scale=.7]{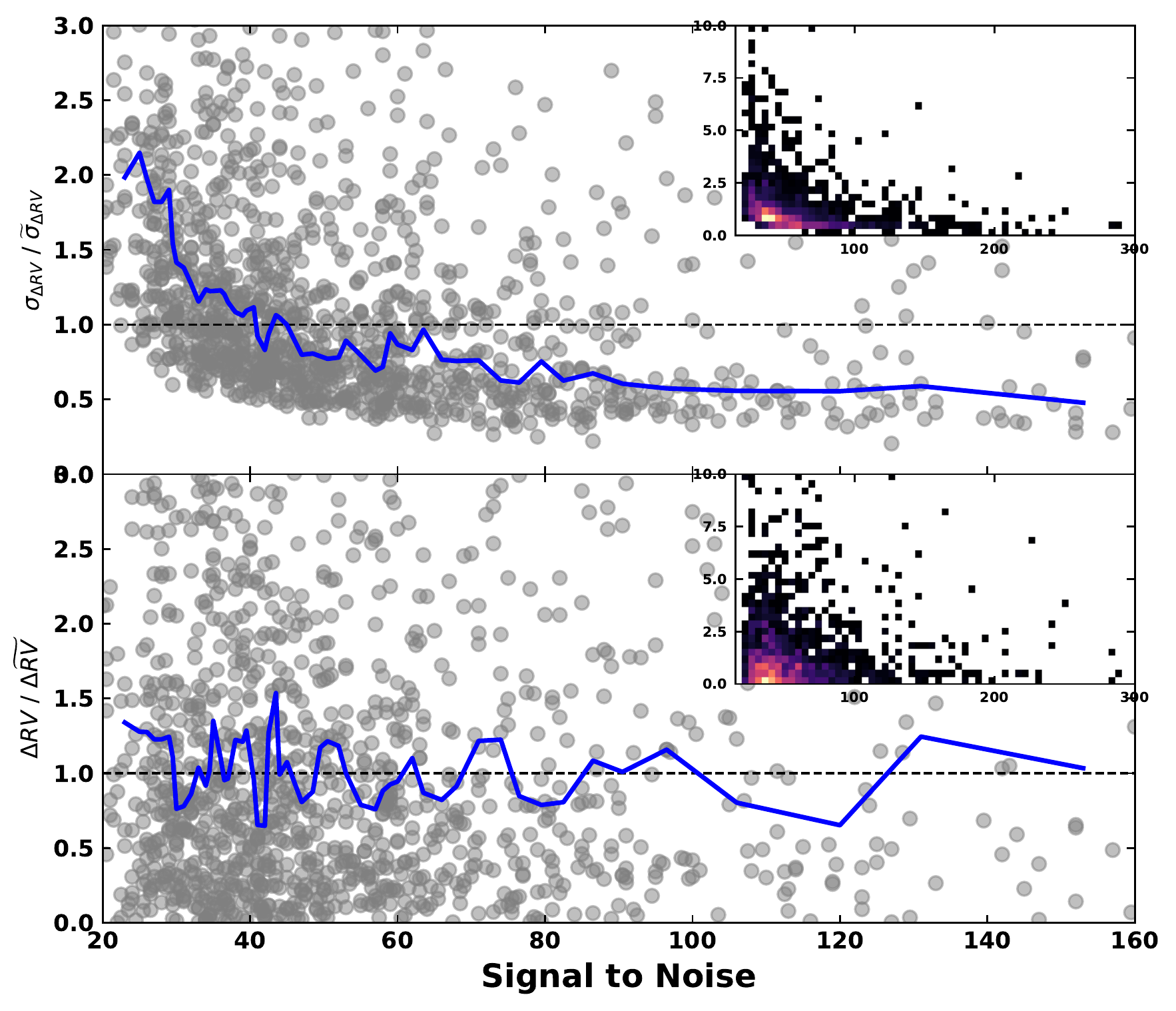}
  \caption{Top Panel: Scatter plot of the \sigdrv\ for our final vetted sample, scaled by the median \sigdrv\, against signal to noise. The blue line is the median overlapping trend line with 50 objects/bin. The horizontal black dashed line demarcates unity with the median \sigdrv\ value. The inset in the upper right shows a density plot of the overall sample space. Bottom Panel: Scatter plot of the \drv\ for our final vetted sample, scaled by the median \drv\, against signal to noise. The blue line is the median overlapping trend line with 50 objects/bin. The horizontal black dashed line demarcates unity with the median \drv\ value. The inset in the upper right shows a density plot of the overall sample space.}
  \label{sn_comp}
\end{center}
\end{figure*}

The \ndrv\ statistic is a ratio of the maximum radial velocity variation within a set of measurements (\drv), weighted by the quadrature error of both measurements on radial velocity(\sigdrv). It is thus important to ensure that there are no inconsistencies between the behavior of either \drv\ or \sigdrv. We checked the \ndrv\ values for our final vetted sample against the derived stellar parameters, \teff\ , \vsini\ , \logg\ , as well as SN. We find no inconsistent behavior between \drv\ or \sigdrv\ when comparing the two to the derived stellar parameters. We do find that the \ndrv\ value went down as we approached smaller SN values. We then checked both \drv\ and \sigdrv, each scaled by the median value of \drv\ and \sigdrv, respectively, against SN. We plot both these comparisons in Figure~\ref{sn_comp}. 

Looking at the bottom panel of Figure~\ref{sn_comp}, we find that there is a relatively flat correlation between the scaled \drv\ and SN, which means any trend to be found between \ndrv\ and SN originates from \sigdrv. From \cite{IN-SYNC_paper_1}, we know that the IN-SYNC pipeline produced stellar parameters errors using signal to noise (see Eqs 3,4, and 5 therein), calibrated by MCMC machinery to ensure that the uncertainty properly matched the epoch-to-epoch variability within stellar parameters. From the top panel in Figure~\ref{sn_comp}, we see that the scaled \sigdrv\ values are indeed correlated with SN at the low values of SN. This suggests that the IN-SYNC pipeline was likely more conservative than warranted at low SN for \sigrv. Checking the scaled per epoch radial velocity errors, we find a similar correlation with signal to noise. 

We proceeded to fit a univariate spline to the scaled per epoch radial velocity errors and signal to noise. Using the resulting function from this fit, we corrected the individual \sigrv\ values using their corresponding signal to noise, and recalculated the \ndrv\ values to ensure that both \drv, and \sigdrv\ behave consistently flat across the signal to noise range of our final vetted sample. 

\kojdel{\textbf{We do note that as the fitted spline corrects the \sigdrv\ values across the entire range of SN, at the higher SN we find an increase in \sigdrv. Though the increase in \sigdrv\ is minor, which could affect our results, we find it a necessary step in order to have consistent \ndrv\ behavior across the entire SN range.}}

\subsection{Robustness of NDRV Statistic Against Stellar Parameters}
\label{ssec:stellar_systematics}
\par
Having accounted for drivers of radial velocity variability resulting from poor data quality, as well as removing stars that have physical characteristics that produce spurious radial velocity variability, we are confident in the capability of the \ndrv\ statistic to detect radial velocity variability arising solely from an unseen companion. As a final argument, we check the \ndrv\ values of our final vetted sample against the derived stellar parameters.

\begin{figure*}[!ht]
\epsscale{1.1}
\plotone{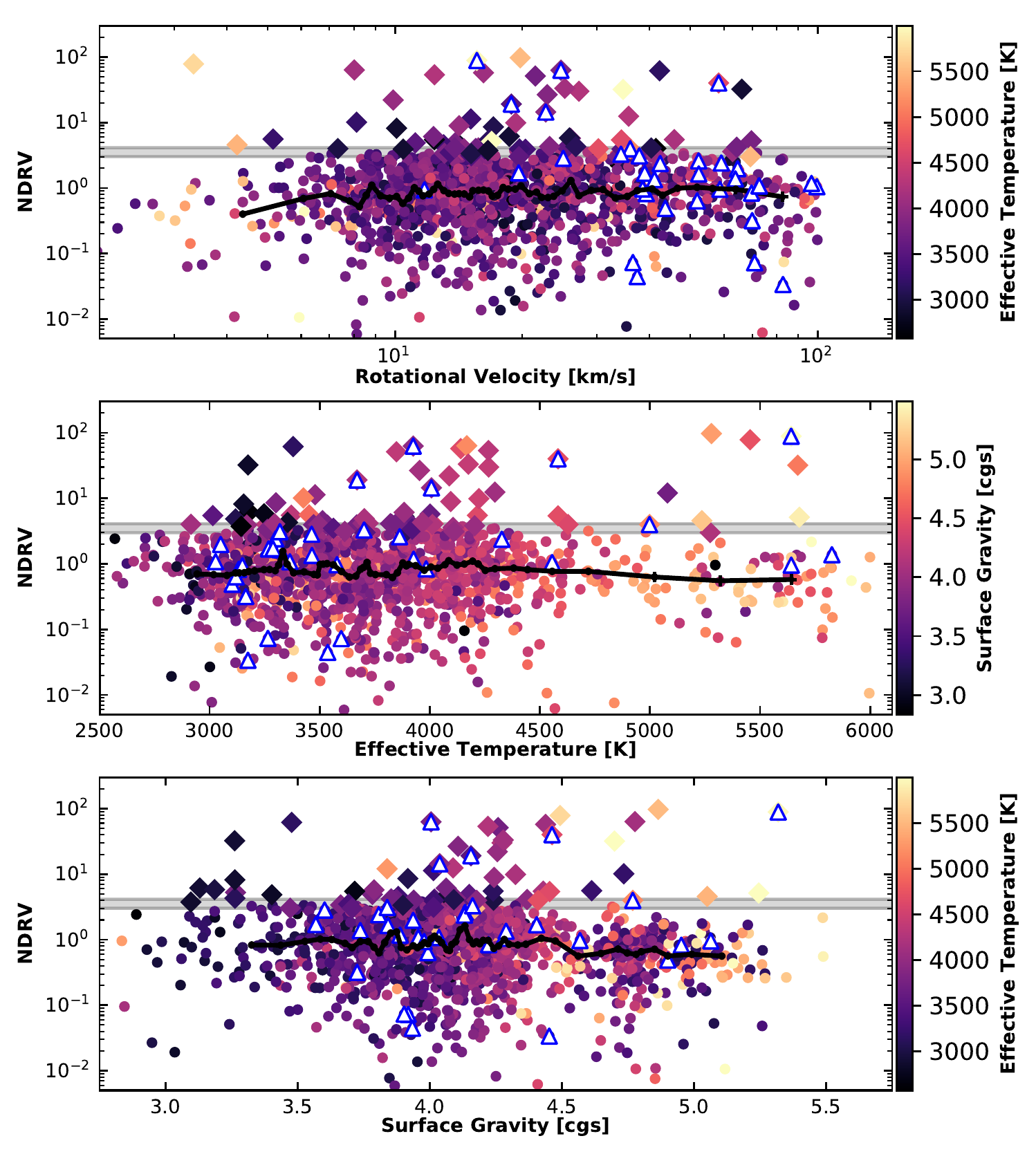}
\parbox{18cm}{\caption{Scatter plot of the Normalized Delta RV$_{max}$ values against the rotational velocity (top panel), effective temperature (middle panel), and surface gravity (bottom panel) values for each star observed in the 5 IN-SYNC clusters and the Pleiades, colored by either \teff\ or \logg. The diamonds are those stars flagged as binary detections using the functional \ndrv\ value, while the circles are stars that were flagged as non-binaries. The gray shaded area is the variable region of $3\sigma$ NDRV values. The lower and upper bounds of the gray shaded area are the $3\sigma$ NDRV values at 2epochs of observation and at 13epochs of observation, respectively, as described in Section~\ref{ssec:ndrv_behavior}. The black trend line in the top, middle, and bottom panel shows the median \ndrv value as a function of half over lapping bins of \vsini, \teff, and \logg\ with 50 objects per bin, respectively. The white triangles with blue outlines are sources within our final vetted that were found to be SB2s within \cite{fernandez2017}.}} 
\label{ndrv_scatter_all}
\end{figure*}

We looked at the behavior of the \ndrv\ statistic as a function of other stellar parameters that should not, in principle, be related to intrinsic radial-velocity variations, in order to see if any biases still existed in the \ndrv\ as a function of the parameters. We calculate the median \ndrv\ values within equal number, half-overlapping bins of the rotational velocity, surface gravity, and effective temperature in order to investigate any potential systematic trends, and plot the \ndrv\ against the aforementioned parameters in Figure~\ref{ndrv_scatter_all}. We find that the \ndrv\ statistic does not significantly correlate with any of the three stellar parameters. 

\kojdel{Figure~\ref{vrad_rms_vsini} shows that indeed the \ndrv\ exhibits significant correlation with the \vsini  \ above $\sim$100~\kms, and with the \teff \ above $\sim$6000~K. We recognize that these stars are F-type stars, which are known for their fast rotation during their pre-main sequence stage.

Thus in our analyses that follow, we have opted to eliminate all stars with a median \vsini \ and/or \teff \ measurement above these limits. We remove 39 stars with 275 observations from our final parent sample with these two constraints.

 Recall that we have already accounted above for the expected effect of rapid rotation (generally among hot stars) on the radial velocity measurements by excluding stars with \vsini \ $ > 100$ ~\kms and/or \teff $ > 6000$~K.}

\clearpage

\subsection{Establishing an NDRV Threshold for Detecting Spectroscopic Binaries}
\label{ssec:ndrv_behavior}
\kojdel{We select spectroscopic binary candidates as those with  \ndrv\ $\geq$ 3.0, i.e., $\Delta RV_{max}$ is at least $3$ times greater than the propagated velocity error.} 

We wish to select binaries as $3\sigma$ detections using the \ndrv\ statistic. We cannot assume a static threshold for all the stars within the IN-SYNC final vetted sample due to the different number of epochs of observation carried out during the IN-SYNC survey (see Figure~\ref{time_baseline_nepochs}). This is due to the $\Delta$RV term (equation~\ref{eq:ndrv}), which measures the greatest variation between any pair of radial velocity measurements within the per epoch data for any one star, increasing as the total number of observations on a star increases as well. The correlation  between the number of RV measurements and the measured $\Delta$RV was previously described in \cite{moaz_badenes2012}.

To establish the $3\sigma$ \ndrv\ threshold, we measure the false positive rate by randomly sampling the RV distribution of a constant RV star with some measurement uncertainty. This distribution is a simple Gaussian with which we generate $10^7$ realizations of randomly sampled pairs, representing two epochs, of RV measurements and calculate their NDRV according to equation~\ref{eq:ndrv}. We then find the \ndrv\  value corresponding to the $3\sigma$ percentile of the distribution. We repeat this process for the the range of the number of epochs ($2$ to $13$) observed in our final vetted sample. We find that the $3\sigma$ threshold monotonically increases with the number of epochs observed. This is as expected; the increased number of samples per realization will increase the expectation value of $\Delta$RV.

We find that the \ndrv\ $3\sigma$ threshold ranges from $\sim$3.00 for 2 epochs of observation up to $\sim$4.11 for 13 epochs of observation. We list the complete list of \ndrv\ $3\sigma$ threshold values in Table~\ref{tab:ndrv_limits}. To characterize binary systems within the IN-SYNC final vetted sample we employ a functional form of the NDRV statistic by comparing any star's \ndrv\ value against the 3$\sigma$ threshold value given how many observations that star had in total within the final vetted sample. Candidate binary systems must have an \ndrv\ value greater than the \ndrv\ value at which there is at least 3$\sigma$ significance. 

\jcbdel{We generate distributions of NDRV values from a single star with a gaussian radial velocity distribution by performing 1e7 samplings with the number of values sampled increasing from 2 up to 13, and then find the NDRV value for each distribution that is equivalent to a 3$\sigma$ value.} 

\begin{figure*}[!ht]
\plotone{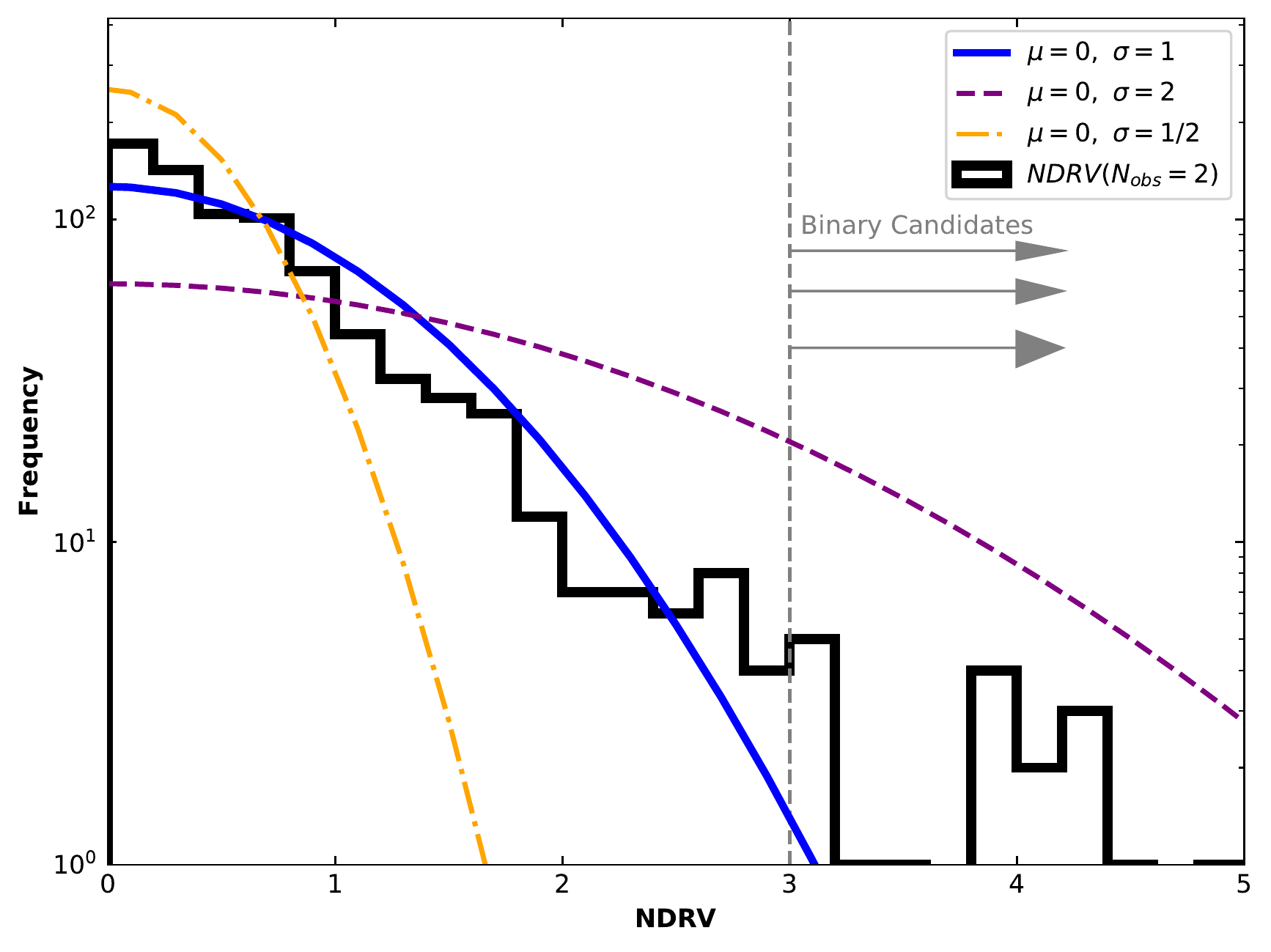}
\parbox{18cm}{\caption{Cumulative distribution of the \ndrv values for the final vetted sample of stars with only two observations within the 6 IN-SYNC clusters, denoted by the black solid line. The solid blue line is a folded normal distribution with $\mu$=0 and $\sigma$=1. The purple dashed line is a folded normal distribution with $\mu$=0 and $\sigma$=2. The orange dotted-dashed line is a folded normal distribution with $\mu$=0 and $\sigma$=.5.}} 
\label{ndrv_distribution_hist}
\end{figure*}

\kojdel{Though a significant fraction of all
stars are likely in binary systems (Section~\ref{sec:intro}), relatively few stars ($\approx7\%$) exceed the NDRV$>3$ threshold due to RV resolution limits and observational properties, both of which we later take into account (Section~\ref{sec:binary_fraction_prob_curve}). Still, the NDRV  distribution qualitatively suggests NDRV$=3$ lies between two major populations: a large group of stars concentrated at low NDRV, presumably single stars or undetected binaries, and a tail of binary systems extending towards high NDRV values (Figure ~\ref{ndrv_distribution_hist}).}

\begin{table}[H]
\begin{center}
\begin{tabular}{| c | c | c |}
  \hline
  $ N_{stars}$ & $N_{epochs}$ & $ 3\sigma\ \ndrv\ Threshold $ \\ \hline
792 & 2 & 3.00 \\ 
310 & 3 & 3.31\\
74 & 4 & 3.49\\
60 & 5 & 3.62\\
72 & 6 & 3.72\\
27 & 7 & 3.81\\
6 & 8 & 3.88\\
13 & 9 & 3.94\\
12 & 10 & 4.04\\
18 & 11 & 4.03\\
14 & 12 & 4.08\\
20 & 13 & 4.11\\
  \hline
\end{tabular}
\caption{\ndrv\ threshold values at $3\sigma$ significance based on the number of epochs from which the \ndrv\ value was calculated. Values listed in second column calculated from Monte Carlo simulations. \label{tab:ndrv_limits}}
\end{center}
\end{table}

The \ndrv\ statistic depends on accurate reporting of the RV measurement uncertainty. We can use our characterization of the false positive rate to assess the accuracy of the reported RV measurement errors. As above, random measurement errors will lead to non-zero NDRV values due to fluctuating radial velocity measurements. Since NDRV is normalized by the measurement errors, the width of the NDRV distribution for non-RV variable objects will  show if the RV errors are under or over estimated. 
\kojdel{Since NDRV is normalized by the measurement error, the distribution of NDRV due to random error should repeat measurements of single stars with non-zero radial velocity errors will produce non-zero NDRV values}
\kojdel{The distribution of NDRV qualitatively suggests NDRV$=3$ lies between  two major components: a large group of stars concentrated at low NDRV, presumably single stars or undetected binaries, and a tail of binary systems extending towards high NDRV values (Figure ~\ref{ndrv_distribution_hist}). We can validate our choice of NDRV threshold and assess the accuracy of the reported RV measurement errors via the detailed shape of the NDRV distribution.}

In Figure~\ref{ndrv_distribution_hist}, we show the overall distribution of \ndrv\ values for all stars with only 2 epochs within the IN-SYNC sample. Unless the RV uncertainties are vastly overestimated, stars with \ndrv$< 3$ can safely be assumed to be non-binaries. For \ndrv\ $\lesssim 3$, the distribution closely follows a folded normal distribution with $\mu=0$, and $\sigma=1$. The same folded normal distribution is an excellent fit to the distribution of $10^7$ \ndrv\ values calculated to determine the false positive rate using two measurement epochs. About $\sim55\%$ of the stars in our final vetted sample have just two epochs. The strong similarity of the \ndrv\ distribution for non-binaries and the folded normal distribution with $\sigma=1$ (solid blue line in Figure~\ref{ndrv_distribution_hist}) indicates that \ndrv\ values for our sample have been calculated with properly estimated errors. If the radial velocity errors were underestimated by a factor of two, the \ndrv\ distribution would broaden by a factor of $2$ (dashed purple line); likewise, if the radial velocity errors were overestimated by a factor of two, the \ndrv\ distribution would broaden by a factor of $1/2$ (dotted-dashed orange line). We conclude that the reported RV uncertainties are accurate. In addition, there are more stars with \ndrv$ \gtrsim 3$ than expected if all stars in our sample were solitary; as demonstrated, these stars are likely to be in binary systems.
 
\subsection{Vetting the NDRV Statistic Against Known Spectroscopic Binary Samples}\label{ssec:ndrv_vetting}

The IN-SYNC derived radial velocity measurements assume that each spectrum eminantes from a single star. It is unclear how this assumption and the NDRV statistic itself affect detection of double lined spectroscopic binaries (SB2s). We now crossmatch our catalog with established spectroscopic binary samples to both characterize how SB2s appear within our analysis framework and validate the recovery of single lined systems (SB1s) using the NDRV statistic.

\cite{fernandez2017} used the APOGEE spectral cross-correlation function to identify 104 SB2 systems within the IN-SYNC survey footprint. Our final vetted sample contains 34 of these systems; the remaining 70 candidate SB2s do not appear in our sample because they either have only one measurement epoch or do not meet are data quality criteria (see Sec.~\ref{ssec:parameters}). Despite using the same the APOGEE / IN-SYNC spectra as \cite{fernandez2017}, we detect only 6 of the 34 crossmatched systems as RV variable using \ndrv. This low SB2 detection rate is perhaps unsurprising given that the IN-SYNC pipeline was not designed to extract parameters from SB2 systems. 

Still, it does imply that our inferred spectroscopic binary fractions for each cluster (Section~\ref{sec:binary_fraction_prob_curve}) underestimates the true spectroscopic binary fraction. However, the absolute binary fraction is not critical to our analysis.  
We are more concerned with the relative cluster-to-cluster binary fractions. Given the distribution of stellar parameters in each cluster (Figure.~\ref{stellar_parameters_edfs}), a correlation between SB2 detection with photospheric parameters could influence even relative measurements. The full 34 known SB2 systems do not exhibit qualitatively different spectral parameters from the rest of the sample (white triangle with blue outlines denote SB2 systems in Fig.~\ref{ndrv_scatter_all}). Splitting the SB2 systems into NDRV-detected and non-detected subgroups, we find no statistic difference in their stellar parameter distributions. We calculate a 2-sided KS statistic of $\sim0.3$ with two-tailed p-values of 0.49, 0.45, and 0.65 for \teff, \vsini, and \logg, respectively. Since there is no significant statistical difference in the properties of SB2 systems that we do and do not detect, we conclude that we can fairly compare the cluster \ndrv\ derived SB fractions.

We proceed to validate the fidelity of our method to recover previously identified SB1s, the assumed binary type in our analysis. \cite{kounkel2016} (hereafter K16) measured radial velocities for 2057 stars in the ONC and NGC 2264, identifying 130 sources as RV variable. Of these 130 RV variable sources, 17 stars appear in our final vetted sample.
However, none of these stars were detected as RV variable according to the \ndrv\ statistic using the IN-SYNC observations.
Starting with our sample of RV variable stars, we identify 11 stars that also appear in the K16 catalog. All 11 of these systems were designated as single stars in K16.
The disagreement in RV variable classification between our two studies is not a failure of our method to recover SB1 systems, but likely the result of differences in the observations of the two surveys. For the 17 RV variable systems only detected in K16, both the median number of epochs (5 in K16; 2 in IN-SYNC) and time baselines ($\sim1100\ \mathrm{d}$ in K16; $\sim11\ \mathrm{d}$ in IN-SYNC) strongly favored RV variability detection in K16. Conversely, the IN-SYNC data favored variability detection (median number of epochs 4.5 in IN-SYNC; 3 in K16) for the 11 systems that only we designate as RV variable.

As the observational properties of each survey strongly affect detection probability of an individual system, we also test whether the \ndrv\ statistic can re-identify the K16 RV variable stars using the K16 data.
We first remove 28 SB2 stars from the K16 RV variable sample and calculate the \ndrv\ of the remaining 102 SB1s using the RV measurements and errors reported in K16. We recover $94\%$ of the K16 variable stars; 96 of the 102 systems had NDRV values exceeding our variability threshold. This successful comparison strongly suggests that the \ndrv\ statistic is a robust method for SB1 detection and gives us confidence to apply the NDRV method to make fair comparisons of the cluster SB fractions using IN-SYNC observations.




\section{\jcbins{Developing Binary Fraction Probability Functions}} \label{sec:binary_fraction_prob_curve}

\subsection{Raw Binary Fractions}

Comparing the \ndrv\ of each star with their corresponding threshold \fndrv\ values, we determine the number of spectroscopic binary (SB) counts for each cluster region, with 10 SB systems out of a total of 88 systems in NGC1333, 19 SB systems out of a total of 455 systems in the Orion A(N) cluster, 11 SB systems out of a total of 312 systems in the Orion A(S) cluster, 7 SB systems out of a total of 110 systems in NGC2264, 22 SB systems out of a total of 237 systems in IC348, and 3 SB system out of a total of 216 systems in the Pleiades. These values are listed in Table \ref{tab:bin_frac}. We also list the 2MASS IDs, \ndrv\ values, and positions of the spectroscopic binary candidates in Table \ref{tab:candidates}.

\subsection{Bayesian Inference Approach} \label{ssec:bayes_infer}

In reality, the raw spectroscopic binary fractions can not be directly compared because of the differing IN-SYNC observational cadences, time baselines, and visitation coverage of the different clusters (Figure~\ref{time_baseline_nepochs}). We have already demonstrated that observational differences between surveys affect RV variability detection (Section~\ref{ssec:ndrv_vetting}). Cluster to cluster differences in observations must be accounted for to ensure a fair comparison of spectroscopic binary fractions across the IN-SYNC survey.  

We therefore develop a probabilistic model of the binary fraction in each cluster to account for the differences in observational data. In addition, we wish to know how the samples we have observed in the IN-SYNC survey relate to the intrinsic populations that exist in these clusters. Our probabilistic approach also accounts for this sampling uncertainty. To constrain the cluster binary fractions, we must calculate the posterior probability $P( X_{f} | O , I )$, where $X_f$ is the binary fraction, $O$ represent the detected binary systems, and $I$ the relevant background information for the data used to determine $O$. Using Bayes' theorem, the posterior probability distribution is proportional to the likelihood of our detected binary systems and any prior on the binary fraction itself

\begin{equation}\label{equation3}
P( X_{f} | O , I ) \propto P( O | X_{f} , I) \times P( X_{f} | I)
\end{equation}
.
\cite{clark2012}  (Hereafter C12) successfully employed a similar framework to construct the binary fraction probability distribution of close period M-type Dwarfs from SDSS data.

We now define the likelihood $P(O| X_f, I)$. Each cluster has $N$ confirmed members of which the subset $O$ are defined as RV variable at the $3\sigma$ level using the \ndrv\ statistic (Equation ~\ref{eq:ndrv} and Section~\ref{ssec:ndrv_behavior}). Since the \fndrv\ measurement of any star is independent of all other stars, \LH\ is the probability of detecting RV variability for each star in $O$ while finding no RV variability for all other cluster stars. The likelihood of assigning the stars to one of these two mutual exclusive groups given the data and a binary fraction $X_f$ is the joint probability of each star's velocity measurements $\{v\}_i$ producing an NDRV value indicative of RV variability ($i \in O$) or not ($i \notin O$):

\begin{equation}
P( O | X_{f}, I ) = \prod_{i \in O}{P( \{v\}_i | X_f, I)} \times \prod_{i \notin O}{P( \{v\}_{i} | X_f, I)}
\end{equation}

where the products are done over all stars in each cluster classified as spectroscopic binaries ($i \in O$) and classified as not spectroscopic binaries ($i \notin O$) using \ndrv.

We are confident that the RV variability quantified by the \fndrv\ statistic is due to the interaction of the primary star with an unseen companion. While this is already a common assertion in the literature \cite[\eg,][C12]{maxted_jeffries2005}, we have demonstrated that neither  spectroscopic parameters (Section ~\ref{ssec:stellar_systematics}) nor RV measurement uncertainties (Section~\ref{ssec:ndrv_behavior}) drive the \fndrv\ statistic .

Therefore, the probability of any system $i$ being detected as a binary is the sum of detection probability given the true binary fraction and the false-positive probability
\begin{equation}
P(\{v\}_i | X_f, I) = X_f \times \pdetecti + (1 - X_f) \times 10^{-2.57} 
\label{eq:pvi}
\end{equation}
where the factor $10^{-2.57}$ is the false positive rate of a $3\sigma$ detection of a binary candidate using \ndrv. The \pdetecti\ term is the probability of detecting a star $i$ as a binary (\ndrv$ > $ \fndrv at $3\sigma$) assuming it is part of a binary system. In the next section, we employ simulated observations of each system to calculate \pdetecti. Since each star is classified as either a binary ($\in O$) or not ($\notin O$), it follows that the probability of non-detection is $1 - P(\{v\}_i | X_f, I)$. 

The final term of the binary fraction posterior probability  is the prior distribution, $P(X_{f}|I)$. This is simply the probability of the model we have chosen given the information we have on hand. Like C12, we also follow \cite{allen2007} in assuming an uninformative prior. We treat the binary fraction as a scale parameter since we are concerned \textbf{only} with relative differences in the binary fraction between clusters. With this information and following C12 we can use Jeffrey's Prior from \cite{sivia_skilling2006} to compose our prior distribution such that $P(X_{f}|I) \propto \frac{1}{X_{f}}$.

\subsection{Calculating P$_{detect}$ with Monte Carlo Simulations}
\label{ssec:monte_carlo_sims}

We still must characterize the probability of falsely classifying a star as non-binary; conversely, the probability of recovering stars in real binary systems as RV-variable binary candidates (\pdetecti). Stars in binary systems trace out periodic velocity curves due to Keplerian motion about the center of mass. The measured instantaneous radial velocity of the star depends upon the orbital phase at which the measurement took place and the intrinsic properties of the binary system. From \cite{lovis2010} we employ the projected velocity vector equation used to find unseen exo-planets orbiting an observed star. Assuming a circular orbit, on-edge binary system, the equation for the expected $RV$ of the primary simplifies to

\begin{equation}
RV(\theta) = RV_{amplitude} \times \sin(\theta),
\label{eq:rv_theta}
\end{equation}

where $\theta$ is the orbital phase of the primary mass relative to our line of sight. Using Kepler's equations of motion, the amplitude of the radial velocity curve under our orbital assumptions is

\begin{equation}
RV_{amplitude} =\sqrt{ \frac{G M_{2}^2}{a(M_1 + M_2)}},
\label{eq:rv_amp}
\end{equation}
where G is the gravitational constant, $M_1$ ($M_2$) is the mass of the primary (secondary), and $a$ is the semi-major axis of the relative binary orbit.

To calculate \pdetecti, we perform mock observations of the primary star motion predicted by equations ~\ref{eq:rv_theta} and ~\ref{eq:rv_amp} within a suite of simulated binary systems. The cadence of the observations and synthetic measurement errors follow directly from the observational properties of our sample (discussed below). The simulated binary systems themselves result from Monte Carlo sampling of the parameter distributions that affect the radial velocity curve -- the binary separation $a$ and the mass ratio $q$.

The  mass ratio $q\equiv M_2/M_1$ for binary stars is relatively well constrained. Following \cite{allen2007}, we model the probability distribution function of $q$ as a power-law $P(q) \propto q^{1.8}$ for $0.02 < q < 0.5$; $0$ otherwise. We do not test other parameterizations of the mass ratio distribution; C12 found no difference in the inferred short-period ($P < 10~d$) binary fraction of SDSS field stars when varying the index of the \cite{allen2007} power-law.

We perform first order mass derivations for the PMS sample and the MS using both Baraffe PMS isochrones \cite{baraffe2015} and Padova PMS isochrones \cite{bressan2012} in order to set an appropriate primary mass for our simulations that would realistically reflect the populations IN-SYNC observed. Choosing Orion A(N) and the Pleiades for our PMS and MS sample, repsectively, we find a median primary mass in both samples of $\sim0.5 \msun$ which we use to set our primary mass limit on our simulated binaries. The secondary mass follows directly from the mass ratio as the primary mass is kept constant.

\begin{figure*}
\plotone{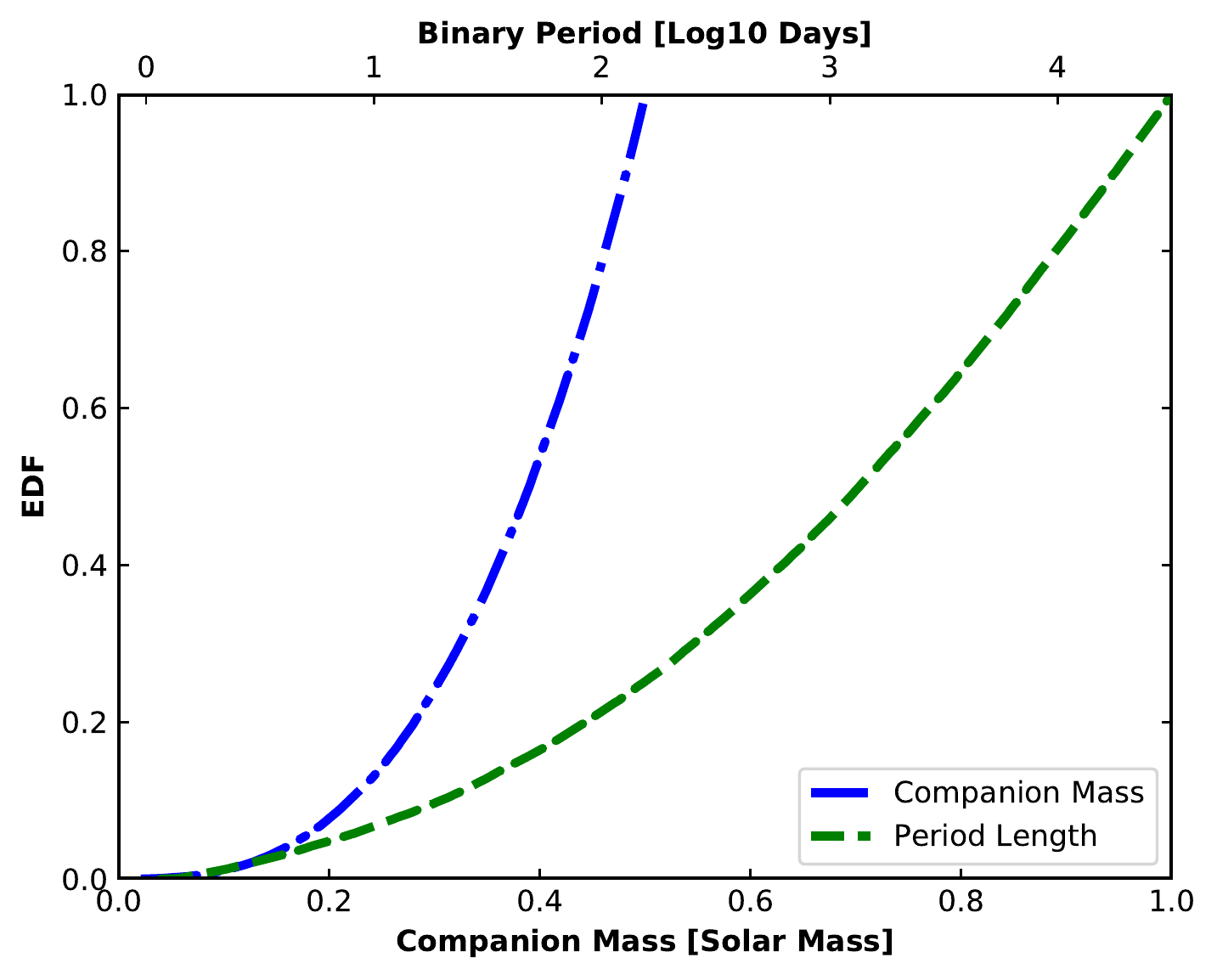}
\parbox{18cm}{\caption{The green dashed line is the empirical distribution function of the periods distribution of binary system simulations used to generate $p_{detect}$ for systems observed in IN-SYNC clusters. Sampled from \cite{DM1991} with a mean of Log10(days) = 4.8, this distribution has been constrained to be below Log10(days) $\le$ 4.5 from tests to calculate the effective period limit at detecting binaries with the IN-SYNC observational parameters. The blue dotted-dashed line is the empirical distribution function of the masses of companions in binary systems simulations to generate recoverability fraction. Sampled from a power law distribution with $\alpha$=1.8 and ranging from 0.02M$_{\odot}$ to 0.5M$_{\odot}$ }}
\label{comb_period_comp_mass}
\end{figure*}

Kepler's Third Law determines the binary separation for stars of known mass given the orbital period of the secondary ($P$). We therefore sample $P$ from an established literature distribution and calculate $a$ for each system via $a^3 = G P^2 (M_1 + M_2) / 4\pi^2$. The distribution of orbital period in our simulated systems is log-normal with mean $\overline{P}= 10^{4.8}$ days and dispersion $\sigma_{P}=10^{2.3}$ days, following \cite{DM1991}. The empirical distribution functions of companion mass and the period distributions that were sampled are plotted in Figure \ref{comb_period_comp_mass}. 

We impose both minimum and maximum period limits on our simulated binary systems. Our minimum period comes from the minimum separation necessary to have a detached, circular 0.5\msun - 0.5\msun \ binary system. For each system composed from randomly sampled companion mass and period, we use the equation for the Roche-Lobe, R$_{l}$, from \cite{eggleton1983}, given below, to ensure that the system is detached,
\begin{equation}
\frac{R_{l}}{a} = \frac{0.49\ q^{-2/3}}{0.6\ q^{-2/3} + \ln(1+q^{-1/3})}.
\end{equation}
We stipulate that the Roche lobe of the primary be less than half of the orbital separation, \ie, $R_l/a < 2$. 

For the maximum orbital period, we recall that the typical RV error is 1.5~\kms, and so even the most advantageous IN-SYNC observational cadence and measurement errors are likely to be unable to recover RV variability at 3$\sigma$ significance for orbital amplitudes $\lesssim 4.5$~\kms. For a 0.5\msun\ binary twin system, this gives a maximum period length of $10^{4.5}$ d.

\begin{figure*}
\begin{center}
\includegraphics[scale=.8]{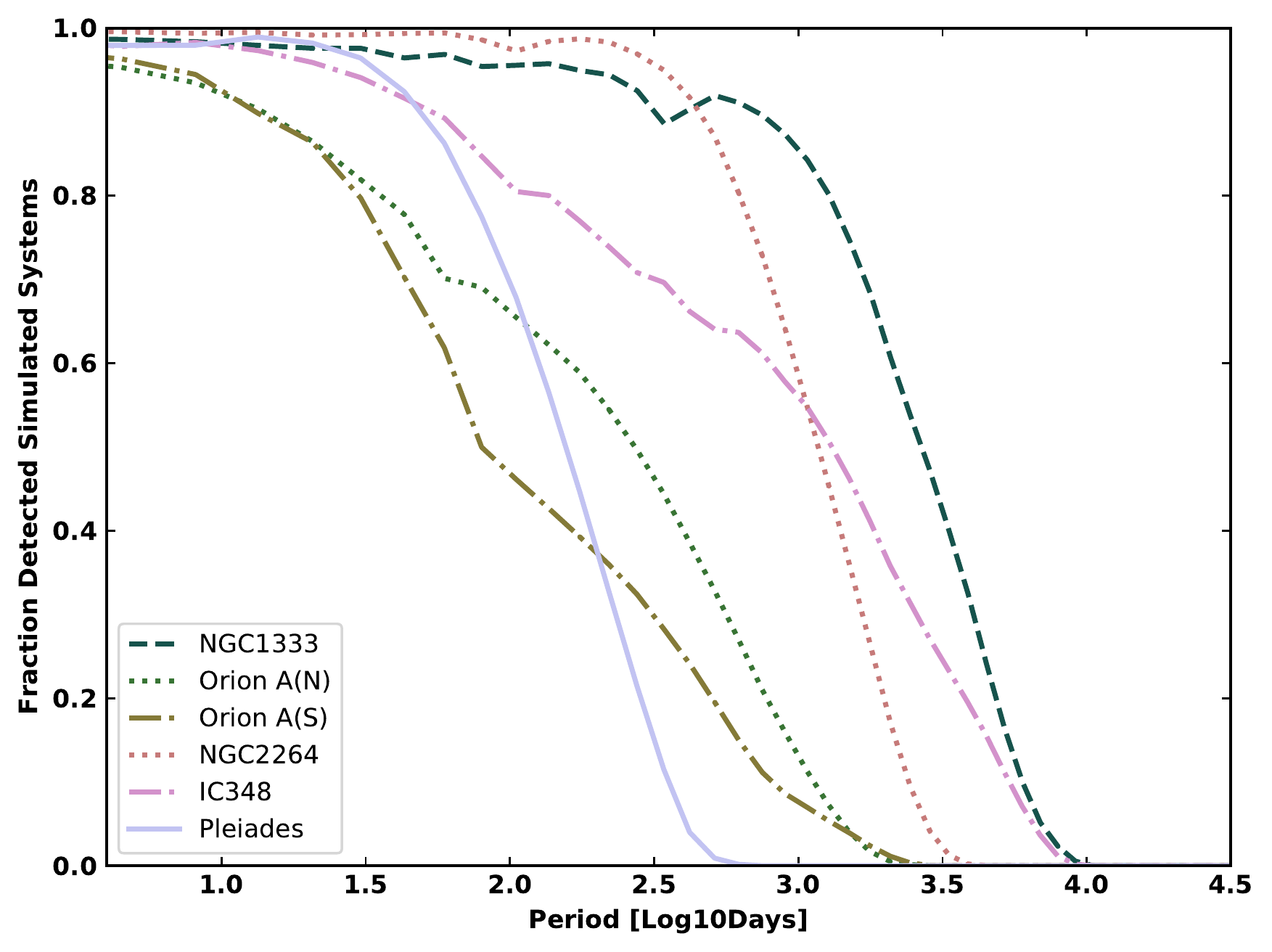}
\parbox{18cm}{\caption{The fractional recovery of simulated systems from Monte Carlo simulations using the observational parameters of the 6 clusters. Each data point represents $\ge$ 1000 simulated binary systems grouped by their period on the x-axis. For each line, there are 1e5 * N$_{cluster}$ simulated systems in total. The y-axis shows the fractional percentage of these systems that produced a $3\sigma$ level \ndrv\ value. As it can be seen here, most of the 6 clusters do not recover any simulated binary systems with the \ndrv statistic with a period of $Log_{10}(days) \ge 4.0$}} 
\label{ndrv_recovery}
 
\includegraphics[scale=1.]{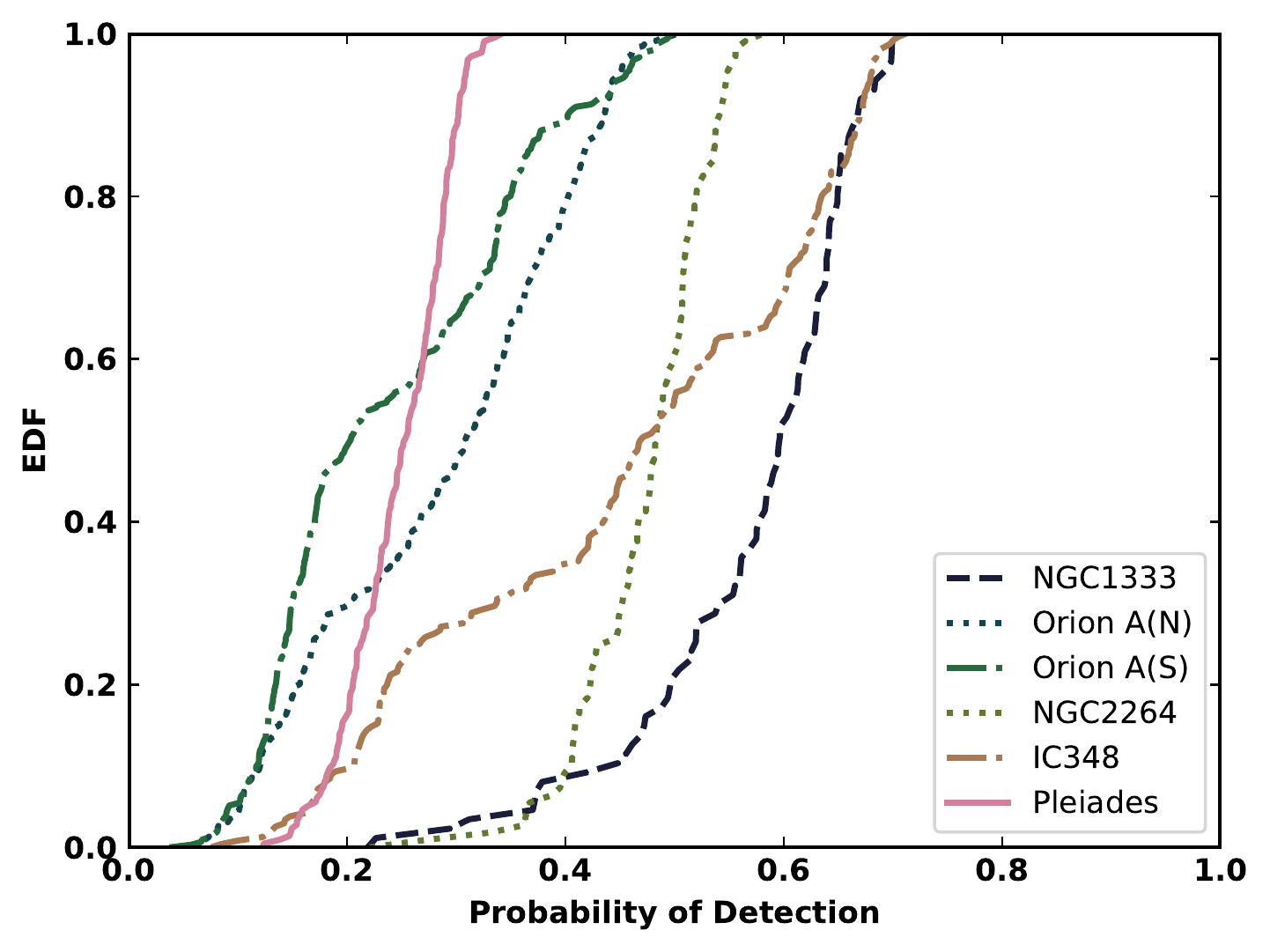}
\parbox{18cm}{\caption{Empirical distribution functions of the $P_{detect}$ values produced from observations of simulated binary star systems produced from Monte Carlo sampling for the 5 IN-SYNC clusters and the Pleiades. }} 
\label{p_detect_dist}
\end{center}
\end{figure*}

To corroborate this choice of maximum period, we calculate the fraction of simulated binary systems recovered using \fndrv\ out of a group of 1e5 simulated orbits for each of our 6 clusters (see Figure \ref{ndrv_recovery}). In each cluster we find that the maximum period recoverable is around $10^{4.1}$~d. We thus set our maximum period cutoff in the simulated orbits to $10^{4.5}$~d, to conservatively assess the recoverability capability of simulated binaries with the IN-SYNC survey. We plot the empirical distribution functions of the probability of detections for each of the 6 clusters in Figure \ref{p_detect_dist}.

It is crucial that the mock observations have the same observational properties as the data. Recall that each star in our final sample has $n = 1 .. N$ ($N \geq 2$) RV measurements with associated errors $\{\sigma_{RV}\}$ collected on $N$ epochs represented by Julian Date and denoted $\{\mathrm{JD}\}$. To observe a simulated system with period $P$, we determine the relative phase $\theta_n$ of our observations, $\theta_n = (\mathrm{JD}_n - \mathrm{JD}_{n-1})/ P$ for $n=1 .. N$; $\theta_0 =0$. The trial radial velocity curve has $N$ RV measurements, computed as $RV_n = RV_{amplitude} \times \sin (\Theta + 2\pi\theta_n)$, where $\Theta$ is a randomly chosen initial orbital phase angle. We then calculate the 
\ndrv\ value of the set of $RV_n$ and the corresponding $\sigma_{RV_n}$ from the actual IN-SYNC observations. For each star and set of observations in our parent sample, we generate $10^5$ binary systems and measure $10^5$ \ndrv\ values. The fraction of these systems with \ndrv $>$ the $3\sigma$ \ndrv\ threshold corresponding to the total number of observations,i.e., tagged as RV variable, is $p_{\mathrm{detect}}$. 

\subsection{The Binary Fraction Posterior Distribution}
\label{ssec:posteriors}
\par
With full expressions for both the likelihood distribution and the prior distribution, the posterior distribution for the binary fraction of a cluster can then be written as follows:
\begin{multline}
p( X_{f} | O , I ) \propto p(O | X_f, I) p(X_f | I) \\
\propto \prod_{i \in O}{p(\{v\}_i | X_f, I)} \times \prod_{i \notin O}{p(\{v\}_i| X_f, I)} \centerdot p(X_f | I) \\
= \bigg[ \prod_{i \in O}{\Big[ X_f \times \pdetecti + (1 - X_f) \times 10^{-2.57}\Big]} \centerdot \\ 
\prod_{i \notin O}{\Big[1 - \big(X_f \times \pdetecti + (1 - X_f) \times 10^{-2.57}\big) \Big]}\bigg] \frac{1}{X_{f}}
\end{multline}

\kojdel{[KARL: OLD equation below
\begin{multline}
p( X_{f} | O , I ) \propto p(O | X_f, I) p(X_f | I) \\
\propto \prod_{i \in O}{p(\{v\}_i | X_f, I)} \times \prod_{i \notin O}{p(\{v\}_i| X_f, I)} \centerdot p(X_f | I) \\
= \bigg[ \prod_{i \in O}{\Big[ X_f \times \pdetecti + (1 - X_f) \times 10^{-2.657}\Big]} \centerdot \\ 
\prod_{i \notin O}{\Big[1 - \big(X_f \times \pdetecti + (1 - X_f) \times 10^{-2.89}\big) \Big]}\bigg] \frac{1}{X_{f}}
\end{multline}
]}

where the products are taken over each star $i$ in the cluster classified as a binary ($i \in O$) or single ($i \notin O$) system.
Finally, we normalize the posterior probability distribution by requiring $\int_{0}^{1} P( X_{f} | O , I )  dX_{f} = 1$.

\section{results}
\label{sec:results}
Applying the methodology laid out in Section~\ref{sec:binary_fraction_prob_curve} to the IN-SYNC observations described in Section~\ref{sec:data}, we obtain the primary results of this study: the posterior distributions of spectroscopic binary frequency for each of the six clusters in our study.

We calculate $p(X_f| O, I)$ for each cluster for $X_f$ in the range of 0 to 1. We also report the median binary fraction within each posterior probability distribution as well as the difference between the median and the 16th and 84th percentiles.  These values are also in Table \ref{tab:bin_frac} along with the number of stars flagged as binary systems using the \ndrv\ statistic, the total number of stars observed within each cluster, the total number of observations within our final vetted data for each cluster, and the age of each cluster from the literature.

\begin{deluxetable*}{ccccccccc}
\tablecaption{In the first 3 columns above we have for each cluster: the 16th percentile of the posterior distribution, the 50th percentile of the posterior distribution, and the 84th percentile of the posterior distribution. In the last four columns we have the number of real systems flagged as binaries, the total number of systems observed for each cluster, the total number of observations within each cluster, the raw spectroscopic binary fraction of each cluster, and the adopted literature age for the cluster. $^{\dagger}$ Age is in Myr. \label{tab:bin_frac}}

\tablehead{\colhead{Cluster} & \colhead{$X_{f,16}$} & \colhead{$X_{f,50}$} & \colhead{$X_{f,84}$} & \colhead{$N_{SB}$} & \colhead{$N_{tot}$} & \colhead{$N_{obs}$} & \colhead{$f_{SB}$} & \colhead{$Age^{\dagger}$} }  

\startdata
  Orion A(N) & 0.104 & 0.134 & 0.168 & 19 & 455 & 1026 & 0.042 & 1.5\\ 
  Orion A(S) & 0.087 & 0.126 & 0.173  & 11 & 312 & 697 & 0.035 & 2.5\\ 			
  IC 348      &0.157 & 0.196 & 0.239 & 22 & 237 & 1375 & 0.093 & 6.0\\ 
  NGC 1333    & 0.132 & 0.184 & 0.246  & 10 & 88 & 423 & 0.114 &1.0\\ 
  NGC 2264    & 0.079 & 0.121 & 0.174  & 7 & 110 & 555 & 0.064 &3.0 \\ 
  Pre Main-Seq  & 0.139 & 0.158 & 0.178  &  69 & 1202 & 4076 & & $\approx$~5    \\ 
  Pleiades   & 0.0086 & 0.0329 & 0.0698 & 3 & 216 & 566 & 0.014 &115\\ 
\enddata
\end{deluxetable*}

\begin{figure*}
\includegraphics[scale=1.2 ]{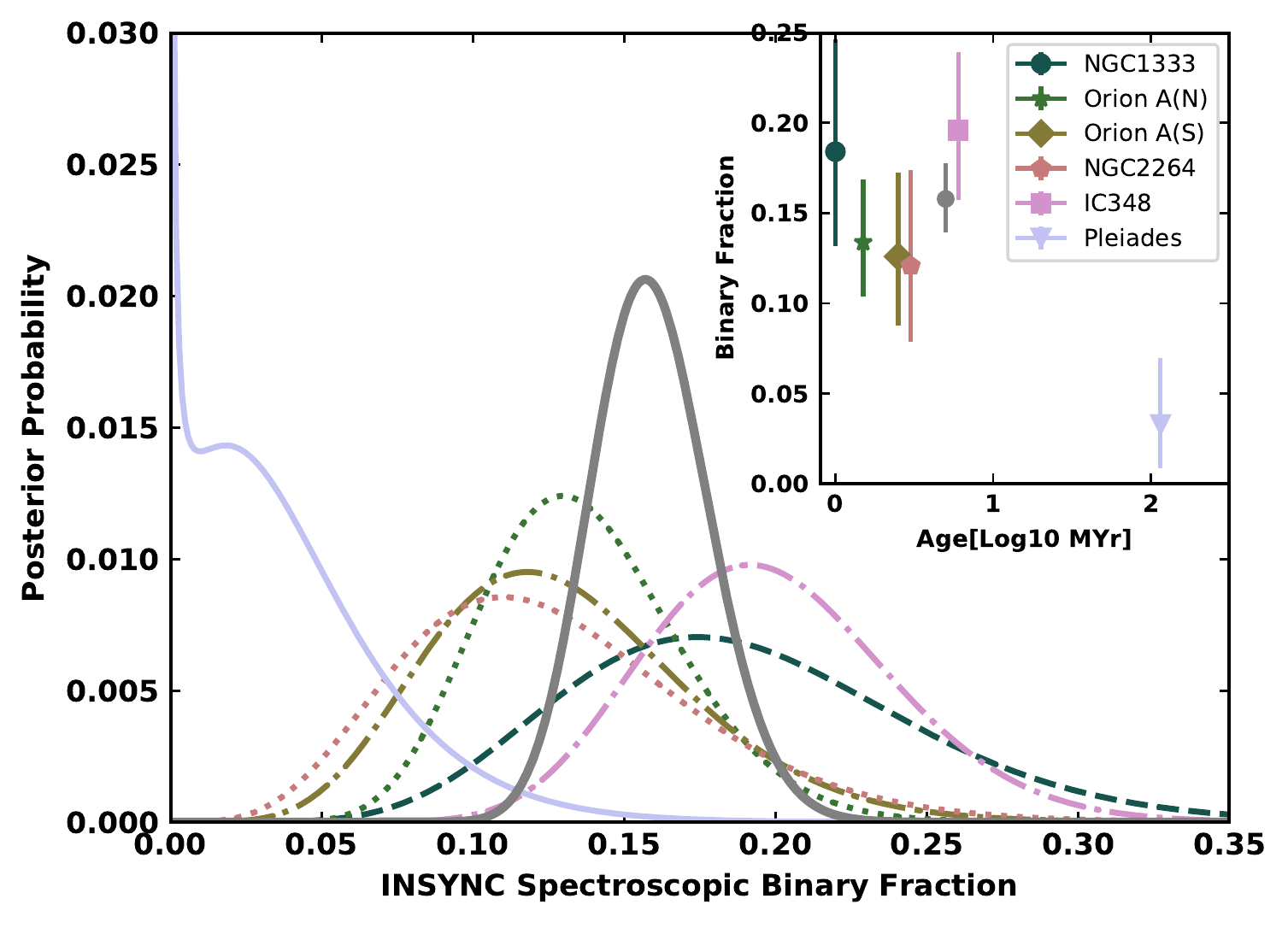}
\parbox{18cm}{\caption{The normalized posterior distributions of the binary fraction for each cluster.  The 5 IN-SYNC clusters (plotted in the dashed lines), and the Pleiades (red solid line).  The gray solid line is the joint distribution of the 5 pre-main sequence IN-SYNC clusters. The inset box in the top right has the ages of the 6 clusters plotted against their median binary fraction, along with their $16^{th}$ and $84^{th}$ percentile values. The joint distribution of the pre-main sequence clusters is also in the inset plot, marked by the gray circle, with $16^{th}$ and $84^{th}$ percentile values. The 'age' of the joint distribution is chosen to be $\sim$5Myr.}} 
\label{prob_bf}
\end{figure*}

Figure \ref{prob_bf} summarizes these key results, where we have plotted the calculated binary fraction posterior distributions for our set of 6 clusters. The inset shows the relationship between the median binary fraction for the 6 clusters and their ages (as discussed in Section~\ref{ssec:data_overview}).  
We see that there is a decrease of the binary fraction from the pre-main sequence ($\sim$1--10~Myr) to the main sequence ($\sim$100~Myr) by a factor of $\sim$ 3--4. 

The shape of the binary fraction posterior distributions of each cluster reflect how the bayesian framework takes into account the number of systems observed overall, the number of systems detected as spectroscopic binaries using the \ndrv\ statistic, and the probability of a binary detection for any star within a cluster. The 5 pre-main sequence clusters have large enough sample sizes, and high enough $p_{detect}$, that the binary fraction posterior distribution can be effectively reconstructed. The Pleiades has the smallest raw binary fraction and the smallest achieved $p_{detect}$ values, which manifests into an asymptotic probability distribution toward the lower binary fraction values. From Figure~\ref{prob_bf} it can be seen that the Pleiades distribution has a turnover occur at the median, after which it becomes asymptotic as it approaches a binary fraction of 0.0. This is likely a result of using Jeffrey's prior, which weights lower binary fraction values as more probabilistically possible given our framework.

The EDFs of the $p_{detects}$ from Figure~\ref{p_detect_dist} reflect how the monte carlo simulations convolve the different time baselines, number of observations (see Figure~\ref{stellar_parameters_edfs}), and real radial velocity error in order to develop the probability of detection to be used within the bayesian framework.

Looking only at the PMS cluster probability distributions, Orion A(N) has the highest peaked cluster. This is most likely due to the cluster having the largest number of systems observed, as well as high values of $p_{detect}$. Comparing with the posterior distribution of IC348, and NGC1333, we find that the the time baselines carried out in each cluster do not play as large a role in constructing the binary fraction distributions as the other factors accounted for. Considering that IC348, and NGC1333 have some of the highest time baselines, as well as the highest $p_{detect}$ values out of the 6 clusters, their resulting binary fraction distribution are more spread out and not as peaked as the two Orion A sub-clusters, suggesting that increasing the number of stars observed will sharpen the resulting distribution, as the intrinsic population will be more comprehensively sampled.

The shape of the Pleiades distribution arises not only from there being only three detected binary within the final 216 stellar systems, but also from the smaller range of detection probabilities that is achieved for the Pleiades cluster. \kojins{For the 5 pre-main sequence clusters, the probability of detection ranges from $\sim$10\% -- 25\% at the lower end up to $\sim$50\% -- 75\% at the upper end. The Pleiades cluster, however, only achieves probabilities of detection ranging from $\sim$15\% to $\sim$35\%.} The shape of the EDFs of the probability of detection is also reflected within the shape of the fraction of detected simulated systems for each cluster. We see that smoothly varying $p_{detect}$ distributions result in smoothly varying behavior in the fractional recovery of simulated binary systems.

The longer right tail of the Pleiades is most likely due to the small number of \ndrv\ binaries found within the cluster sample, and the values of probability of detection achieved. Since the $p_{detect}$ values were the smallest out of the 6 clusters observed by IN-SYNC, there is a higher possibility that some of the stars designated by the \ndrv\ statistic as non-binaries were indeed false-negatives, and suggests that the true binary fraction of the Pleiades could be higher.

We can use these probability distributions to assess the statistical significance of the our finding of a declining binary fraction. Considering for example just the pairwise comparison of the Orion A North region versus the Pleiades, and using a simple $\chi^2$ null hypothesis test, we conclude that the difference is modestly significant at $\lesssim$2$\sigma$. However, the same difference in both sense and magnitude remains for each of the clusters, strengthening confidence in the result.

We can also consider the joint probability distribution of all five pre--main-sequence clusters as one. Having a total number of 1202 stars, as well as combining some of the higher distributions of probability of detection, it is the most sharply defined and most highly peaked distribution.  This is shown in Table \ref{tab:bin_frac} and Figure~\ref{prob_bf} (gray solid line). Taking the pair wise comparison of the joint pre-main sequence distribution and the pleiades, the difference from the pre--main-sequence to the main sequence becomes more highly significant at the 3--4$\sigma$ level.

\section{discussion}
\label{sec:discussion}

In this work, we have used simulated observations within a Bayesian framework to infer the spectroscopic binary fractions of young star clusters, based on observations of the pre-main sequence aged IN-SYNC clusters to the main-sequence aged Pleiades cluster. From our simulations of detached, edge-on, circular spectroscopic binaries, using the \ndrv\ statistic and accounting for the IN-SYNC time-baselines, number of observations, and radial velocity errors, we find that the IN-SYNC survey has a sensitivity to spectroscopic binaries with orbital periods ranging from $10^{2}$~d -- $10^{4.1}$~d, depending on the cluster observed. The observed samples are found to all represent a similar mass regime, with a median estimated mass of $\sim$0.5~\msun. Thus, our results apply to low-mass binaries of short- and intermediate-periods.

Our results suggests that the fraction of low-mass binaries with orbital periods up to $10^{4.1}$~d appears to be decreasing by a factor of $\sim$3-4 from $\sim$1~Myr to $\sim$100~Myr.

In the following subsections, we discuss the relevant observational literature to place our findings in context. We then specifically discuss prior theoretical findings regarding the role of dynamics in the evolution of binaries. Finally, we consider potentially important effects that we have not yet included in our analysis.

\subsection{Comparison with previous observational results}\label{ssec:}

We find that there is general agreement between our inferred spectroscopic binary fractions for the IN-SYNC pre-sequence clusters and previous observational studies of pre-main sequence spectroscopic binaries. 
For example, looking at the Ophiuchus region in the H-band, \cite{prato2007} found a spectroscopic binary fraction of $12^{+8}_{-3.5}$. Their sample consisted of 33 T-Tauri stars observed over 3 years with 10m Keck-II Telescope, mostly consisting of M-type stars. Our derived spectroscopic binary fractions for all 5 of our pre-main-sequence regions fall within this observed range.

Similarly, \cite{tobin2009, tobin2013} looked at the Orion Nebula Cluster's spectroscopic binary fraction, gathering multi-epoch data for a larger sample of 727 stars, achieving sensitivity to spectroscopic binaries out to $4\times10^{3}$ days. With this multi-epoch data they find 89 binaries \kgsdelete{within their sample}, giving a spectroscopic binary fraction of 11.5\% for the ONC, which is slightly lower than but comparable to our spectroscopic binary fraction of $\sim$~13.4\%. 

\citet{kounkel2016} did a reanalysis of the \citet{tobin2009} Orion data and also performed an analysis of NGC2264. They found their binary fractions within separations of 10~AU to be $\sim$~5.3\% and $\sim$~5.8\%, respectively. These observed binary fractions are roughly half of the values we have inferred using our Bayesian framework of $\sim$~13.4\% and $\sim$~12.1\% for Orion A(N) and NGC~2264, respectively.
However, the IN-SYNC survey
has a sensitivity out to 20~AU and so we would expect a higher number of binaries to be detected as the parameter space of detection is increased.

\cite{mathieu1989} also found a relatively low pre-main sequence spectroscopic binary fraction. They found 6 young spectroscopic binaries within the naked T-Tauri star populations of Taurus-Auriga, Scorpius-Ophiuchus, and the Corona-Australis star forming regions for short periods (p$<$100 days), corresponding to a spectroscopic binary fraction of $9\pm 4$\%. This again is not necessarily inconsistent with the higher binary fractions found by IN-SYNC and the other studies noted above that probe a much larger range of binary orbital periods.


Next we consider how our inferred spectroscopy binary fraction for the Pleiades compares to previous findings for main-sequence clusters. Our results are most directly comparable to \cite{abt1987} who derived a short period main-sequence binary fraction of 12\%. This is a factor of $\sim$3 higher than our measurement, however
\cite{abt1987} considered stars roughly 2--3 times more massive than those observed by IN-SYNC, and the short-period binary fraction has been found to increase linearly with primary mass \citep[][]{clark2012}.

In contrast, using the CORAVEL radial velocity survey, \cite{raboud_mermilliod1998} (hereafter RM1998) found the Pleiades overall binary fraction to be a higher $\sim$20\% within a mass range from 0.5\msun -- 1.0\msun.
As with \cite{abt1987}, the sample observed by RM1998 covered a higher mass range (see their Figure 10)
In addition \kgsdelete{to this}, the time baseline of the \kgsdelete{spectrocopic CORAVEL data used in} RM1998 observations spanned 17 years, much longer than the $\sim$1.3 year maximum time baseline of the IN-SYNC survey. 
Thus, the binary fraction determined by RM1998 represents both a larger period range and a higher mass range than the IN-SYNC sample, which as noted above is more directly comparable to the \cite{abt1987} results.

\subsection{Dynamical  evolution of short-period versus intermediate-period binaries} \label{ssec:dynamics}

Observationally, there has been evidence for a decreasing overall binary fraction on long timescales among solar-type stars \citep[$\gtrsim$1~Gyr][]{} \citep{mason1998}. During the pre-main-sequence phase, observational results have been sparse, but generally have also suggested a overall decreasing binary fraction from pre-main-sequence ages to field ages \citep[e.g.,][]{ghez1993, ghez1997}. However, these results have typically been based on longer period binary populations, much wider than those accessible to the IN-SYNC observations.

The fact that our results pertain to shorter period spectroscopic binaries could on its face be surprising, considering that simulations have shown short-period binaries in fact become more tightly bound as they evolve within a cluster \kgsdelete{a behavior commonly known as {\it Heggie's Law}} \citep[see][and references therein]{heggie1975}. This may imply that it is specifically the intermediate-period binaries (periods of $\sim$10$^2$--10$^4$~d), which have not been well probed in previous observational studies of pre-main sequence populations, that are undergoing dynamical disruption as they evolve toward Pleiades age. 

It is expected that dynamical processing via interactions with other stars in the cluster will  modify the number of binaries, and the distributions of their properties, over time. Current theory relies largely on N-body simulations to reproduce the observed states of star clusters. 
For example, \cite{kroupa_2001_binaries} finds that the overall number of binaries with period range $10^0$--$10^9$~d of a simulated cluster decreases after just $\sim$2.5~Myr, comparable to the typical age of the IN-SYNC cluster sample, as binary disruption occurs among the widest binaries, unbinding the smaller mass secondary star. \cite{goodwin_kroupa2005} further finds that the disruption is most pronounced for lower-mass M-dwarf binaries due to the lower binding energy.
Thus, the decrease in the spectroscopic binary fraction that we observe between the pre-main sequence clusters and the Pleiades, could be a manifestation of these effects, particularly given the low masses that characterize our observational samples and the sensitivity of our observations to relatively long-period binaries.

At the same time, the role of cluster environment is a potentially confounding factor. \cite{duchene1999} did not find clear evidence for a decrease in the binary fraction on the basis of comparing their IC~348 sample with the Trapezium \citep{prosser1994, petr1998}, the Pleaides \citep{bouvier1997}, and solar-type stars in the field \citep{DM1991}. However, they do postulate that there may be an anti-correlation between binary fraction and cluster density; i.e., that the densest star-forming environments (e.g., Orion Trapezium region) have a lower binary fraction than low-density environments (e.g., Taurus-Auriga). That result would be consistent with the literature on binary fractions in older globular clusters \citep[][but see \citet{milone2012}]{sollima2007}.

More generally, the argument of cluster density driving the the observed differences of the binary fraction among different clusters would complicate the apparent evolution of the binary fraction in our samples if the Pleiades cluster had started as a pre-main sequence cluster with higher stellar density than the IN-SYNC clusters.

\cite{converse_stahler2009} used simulations in an attempt to recreate the initial dynamical state of the Pleiades. Their results suggest that the Pleiades cluster already had a high degree of mass segregation, and expanded rapidly, leaving behind only a dense core remnant. 
\cite{fuente_marcos1997}, \cite{fuente_marcos1998}, and \cite{adams2000} corroborate the scenario that dense older clusters like the Pleiades are the dense bound remnant core of a larger open cluster that has already evaporated away. 

However, \cite{kroupa_2001_sc} suggests that the Pleiades open core remnant observed today could have arisen from a high stellar density OB association such as the ONC. This cluster nucleus would be all that was left after the high mass O stars expelled the gas cluster gas, causing the majority of the cluster to expand outward on shorter relaxation time scales, while the core became more bound due to two-body relaxation avoiding the cluster dissolution. Simulations performed by \cite{kah2001} support this scenario, finding a cluster simulation of the ONC leading to a Pleiades-like remnant core. They find that the binary fraction within this dense OB association would also decrease over the evolution of the cluster to a dense core remnant. 

It is clear that more work is needed to solidify the conclusion that the Pleiades represents the later-stage evolution of pre-main sequence clusters similar in density to those in our IN-SYNC sample. However, assuming that the IN-SYNC clusters and the Pleiades may indeed be fairly compared, we suggest that the most relevant consideration for the present work is that the parameter spaces of previous observational studies had a dearth of observational information on pre-main-sequence binaries with periods ranging from $\sim10^{2}$days -- $10^{5}$days. Considering that the IN-SYNC survey achieved sensitivity over the period range of $10^2$--$10^4$~d, we speculate that the binary fraction decrease found in our results stems from these relatively shorter period binaries which are still wide enough to be disrupted by dynamical interactions over the 100~Myr timescale we  consider here.

\subsection{Caveats and issues not addressed in this work }

We have not attempted to treat the influence of triples and higher-order hierarchal systems. \cite{tokovinin2006}, looking at a sample of 165 solar-type spectroscopic binaries, found that as many as $\sim$80\% possess a tertiary component when the period of the inner binary pair is less than 7~days, decreasing to $\sim$30\% for inner-binary period  $\gtrsim$20~days. While our observed and simulated parameter space spans a very much larger period range, the apparent decrease in binary fraction that we observe could in principle result if the IN-SYNC clusters possess a higher fraction of triples than the Pleiades, thus artificially inflating the incidence of (easier to detect) short-period spectroscopic binaries in our pre-main sequence samples. 

The Bayesian framework that we have developed should be useful for future efforts to infer binary populations from inhomogeneous datasets. Even so, there are also directions for improving our framework. With regard to our simulations, we have ignored the effects of eccentricity and inclination on generating radial velocities to be simulated using the observational time baselines and observational cadences of our 6 clusters. Thus all of the radial velocity curves for our simulated binaries before being `observed' are all symmetric and sinusoidal. 
Fortunately, however, for the purposes of our work, these effects will largely be systematic in nature, but the relative binary fractions should be largely preserved. For example, inclination will systematically decrease the amplitude of the radial velocity curves of the simulated binaries, and thus will lower the $p_{detect}$ values for all 6 clusters, thus underestimating the inferred binary fractions. 

The effect of eccentricity is somewhat more complex. On the one hand, circularization has been found observationally to occur for binaries with a period $\le$10 days \citep{meibom2005, raghavan2010}, and so our assumption of circular orbits in our simulations should be robust for the shortest period spectroscopic binaries. For longer period binaries, there is a competing effect of eccentric binaries being easier to detect spectroscopically near periastron but more difficult to detect near apastron. We expect that the net effect will not be large, but this is an avenue for future extensions of the simulations we have presented here. 

In addition, we have not attempted to model the effects of the increased magnetic activity of pre-main-sequence stars is known to drive spurious RV jitter. \cite{hillenbrand2015} calculated the lower limit of mass for planetary companions around stars younger than our Sun and found RV jitter is correlated with chromospheric activity. They find the highest levels of chromospheric activity produce RV jitter of $\sim$200~\ms. \cite{stassun2004} finds that sunspot activity can cause spurious RV jitter because of contrast between the sunspot and the stellar photosphere. This RV jitter can be of order 1~\kms\ in the optical. We tested what effect this would have upon our results and introduced a sunspot RV jitter error correction into the denominator of our \ndrv\ statistic (see equation \ref{eq:ndrv}) of $\sigma_{sunspot}= 0.25~\kms$. We adopted a value of 0.25~\kms\ for sunspot activity as \cite{marchwinski2015} found that RV jitter in the near IR is up to 4x less than in the optical \cite[see also][]{mahmud2011}. If we assume that this error applies to all of the stars in our pre-main sequence samples, then our finding of a decreasing binary fraction from the pre-main sequence to the Pleiades declines to $\sim2\sigma$ significance.

More generally, our work relies on the measured RVs and RV errors from the IN-SYNC pipeline. In any analysis, the robustness of the measurements is crucial to the robustness of the results derived from them. Throughout this work we have attempted to characterize and account for systematics present within the data we use---across stellar parameter space, within a given cluster, across the various clusters---as well as systematics that may arise from our analysis machinery itself. As with any work, any deeper or hidden systematic effects that we have not considered will of course remain unaccounted for. In this section we have attempted to identify a number of these potential additional systematic effects. We cannot know the extent to which they actually impact our data or analysis, but we may speculate that our estimates above suggest that even if present these effects do not negate the findings of this work.



\section{summary} 
\label{sec:summary}
Using the high precision ($\sigma_{RV}\sim .3\kms$), high volume ($>10^{3}$ stars) H-Band spectroscopic data from the IN-SYNC survey, we studied the derived stellar properties of five IN-SYNC clusters to find signals of possible unseen sub-stellar companions. We used the \fndrv\ statistic to find candidates that could have unseen stellar companions within the IN-SYNC data. The advantage of the \fndrv\ is its ability to capture the maximum $\Delta$RV within a set of radial velocity measurements and weighting it by the relative error of the two measurements that produce the highest $\Delta$RV with as little as two observations of a star.  

We performed Monte Carlo simulations within a Bayesian framework in order to account for the differing observational cadences and other observational biases in the observations from one cluster to the next. 
Due to the small number of systems within each region observed by the IN-SYNC survey as well as the Pleiades, it was not enough to simply bootstrap our data in order to characterize the error on the raw spectroscopic binary fraction. We could not assume that the sample of systems observed in the IN-SYNC survey is representative of the population of star systems within these 5 clusters. In order to characterize the spectroscopic binary fraction of these clusters, we developed a Bayesian framework in order to infer the underlying distributions for each cluster while taking into account the calculated effectiveness of IN-SYNC observations at capturing radial velocity variations from unseen companions using simulated observations of constructed binaries. The framework that we have presented should itself be useful for future studies that seek to make inferences about binary populations from inhomogeneous datasets.

We find the spectroscopic binary fraction for the five pre--main-sequence clusters, with ages in the range $\approx$1--10~Myr, to be in the range $\approx$20--30\%.
Performing the same \fndrv\ analysis on similarly reduced, APOGEE derived, spectroscopic data from the Pleiades, we find a smaller spectroscopic binary fraction of 5--10\%. Even with all of the biases and sampling effects folded into the analysis, this decline in binary fraction is modestly significant when comparing any one of the pre--main-sequence clusters to the Pleiades. That the same decline in sense and in magnitude emerges for each of the pre--main-sequence clusters, bolsters this conclusion, and becomes significant at the 3--4$\sigma$ level when all of the pre--main-sequence clusters are considered together.

There are additional effects, in particular random line-of-sight inclinations and eccentric orbits for long-period binaries, that we have not included here but that would allow our framework to produce inferred binary fractions that are closer to the absolute binary fraction. In addition to this, in this work we considered the effect of RV jitter from pre-main-sequence magnetic activity in a simplistic fashion by assuming that it could negatively affect every star in the sample, which in turn would decrease the statistical significance of our result. This effect could be more fully modeled or treated more probabilistically in future work.

In the meantime, we have focused here on relative differences across the ages of the clusters in our study.
Importantly, the APOGEE observations considered in this work have a sufficient cadence and span a sufficient time baseline to allow us to probe low-mass spectroscopic binaries with orbital periods as long as $\sim$10$^4$~d. Most previous observational studies of binary populations in pre-main sequence clusters have been sensitive to either much tighter spectroscopic binaries and/or to much wider visual binaries. Binaries in the period range $\sim$10$^2$--10$^4$~d have not been well studied by spectroscopic observations heretofore. Our results suggest that these intermediate-period binaries are tight enough to be detectable in the APOGEE radial-velocity measurements yet soft enough to be susceptible to dynamical sculpting in their birth clusters prior to arrival on the main sequence.

\acknowledgements

We would like to thank the referee whose comments greatly helped to improve the quality and scope of this paper. We would also like to thank Dr. Kaitlin Kratter and Dr. Moe Maxwell for fruitful discussions that helped clarify the impact of the results in this paper.

This work was funded by an NSF LSAMP Bridge to the Doctorate grant. K.G.S.\ acknowledges partial support from NSF PAARE grant AST-1358862. Funding for SDSS-III has been provided by the Alfred P. Sloan Foundation, the Participating Institutions, the National Science Foundation, and the U.S. Department of Energy Office of Science. The SDSS-III web site is http://www.sdss3.org/.

SDSS-III is managed by the Astrophysical Research Consortium for the Participating Institutions of the SDSS-III Collaboration including the University of Arizona, the Brazilian Participation Group, Brookhaven National Laboratory, Carnegie Mellon University, University of Florida, the French Participation Group, the German Participation Group, Harvard University, the Instituto de Astrofisica de Canarias, the Michigan State/Notre Dame/JINA Participation Group, Johns Hopkins University, Lawrence Berkeley National Laboratory, Max Planck Institute for Astrophysics, Max Planck Institute for Extraterrestrial Physics, New Mexico State University, New York University, Ohio State University, Pennsylvania State University, University of Portsmouth, Princeton University, the Spanish Participation Group, University of Tokyo, University of Utah, Vanderbilt University, University of Virginia, University of Washington, and Yale University.

\newpage

\begin{deluxetable*}{ccccccc}
\tablecaption{Spectroscopic binary candidates identified within the IN-SYNC final vetted sample using the \ndrv\ statistic. The first three columns give the 2MASS ID for the binary, its cluster membership, and the number of observations for that system within the final vetted sample. The next two columns list the maximum $\Delta_{RV}$ value for that system as well as its \ndrv\ value. The last two columns give the binary's position data in decimal degrees.\label{tab:candidates}}

\tablehead{\colhead{2MASS ID} & \colhead{Cluster} & \colhead{$N_{obs}$} & \colhead{$\Delta_{RV}$~[\kms]} & \colhead{NDRV} & \colhead{$\alpha$~(J2000)} & \colhead{$\delta$~(J2000)}  } 

\startdata
2M03273825+3013585&NGC1333&2&6.08&3.02&51.909394&30.232935 \\
2M03274767+3012043&NGC1333&6&18.24&5.26&51.948666&30.201195 \\
2M03285097+3123479&NGC1333&2&4.33&3.93&52.212378&31.396646 \\
2M03285101+3118184&NGC1333&3&11.98&6.15&52.21256&31.305134 \\
2M03292187+3115363&NGC1333&6&14.3&5.48&52.341137&31.260084 \\
2M03292349+3123309&NGC1333&4&2.98&4.15&52.347904&31.391941 \\
2M03305456+3031466&NGC1333&7&6.15&12.06&52.727365&30.529621 \\
2M03311830+3049395&NGC1333&2&1.62&3.81&52.826288&30.827658 \\
2M03323300+3102216&NGC1333&7&3.63&4.85&53.137523&31.039352 \\
2M03331284+3121241&NGC1333&7&2.93&4.89&53.303514&31.356705 \\
2M03411921+3202037&IC 348&9&3.15&4.56&55.330048&32.03437 \\
2M03415507+3149394&IC 348&3&34.69&78.31&55.479492&31.82762 \\
2M03423215+3229291&IC 348&9&4.77&3.98&55.633987&32.49144 \\
2M03425395+3219279&IC 348&5&6.12&4.02&55.724817&32.324429 \\
2M03430679+3148204&IC 348&10&12.96&19.09&55.778292&31.805687 \\
2M03435208+3203398&IC 348&8&1.75&3.99&55.967013&32.061073 \\
2M03435550+3209321&IC 348&4&65.11&97.23&55.98128&32.158939 \\
2M03440216+3219399&IC 348&12&4.68&4.1&56.009038&32.327766 \\
2M03441297+3201354&IC 348&7&600.85&30&56.054083&32.026527 \\
2M03442228+3205427&IC 348&2&10.7&22.01&56.092865&32.095219 \\
2M03442602+3204304&IC 348&2&5.01&5.42&56.108446&32.075115 \\
2M03442766+3233495&IC 348&11&3.94&4.17&56.115276&32.563755 \\
2M03442851+3159539&IC 348&4&3.03&3.9&56.118804&31.998333 \\
2M03442997+3219227&IC 348&13&9.16&8.57&56.124912&32.32299 \\
2M03443694+3206453&IC 348&2&4.16&3&56.153922&32.112602 \\
2M03443739+3212241&IC 348&2&2.96&4.9&56.155794&32.206722 \\
2M03443798+3203296&IC 348&13&41.59&57.36&56.158251&32.058239 \\
2M03444306+3137338&IC 348&6&2.48&3.93&56.17942&31.626057 \\
2M03445200+3226253&IC 348&11&5.88&5.52&56.216675&32.440369 \\
2M03450326+2350220&Pleiades&2&2.37&4.56&56.2636&23.83944 \\
2M03450762+3210279&IC 348&4&20.75&31.92&56.281784&32.174427 \\
2M03452106+3218178&IC 348&10&11.01&11.32&56.337768&32.304951 \\
2M03454440+2413132&Pleiades&3&58.72&88.71&56.43504&24.220335 \\
2M03473521+2532383&Pleiades&3&22.1&63.4&56.89671&25.54397 \\
2M03474711+3304034&IC 348&13&4&5.75&56.94631&33.067616 \\
2M05333425-0443458&Orion A(N)&2&17.72&26.58&83.392737&-4.72939 \\
2M05340819-0511429&Orion A(N)&2&2.52&5.59&83.534145&-5.19525 \\
2M05341723-0529043&Orion A(N)&2&4.58&5.81&83.571796&-5.484531 \\
2M05342829-0610247&Orion A(S)&4&18.18&12.44&83.617885&-6.173548 \\
2M05343487-0611259&Orion A(S)&2&147.63&32.28&83.645318&-6.190552 \\
2M05344830-0537228&Orion A(N)&3&5.76&3.6&83.701287&-5.623006 \\
2M05345259-0515366&Orion A(N)&2&4.75&3.11&83.719149&-5.260176 \\
2M05350200-0520550&Orion A(N)&2&18.38&14.45&83.758339&-5.348637 \\
2M05350392-0529033&Orion A(N)&3&85.01&62.68&83.766341&-5.484265 \\
2M05350560-0518248&Orion A(N)&2&3.88&3.95&83.77337&-5.306904 \\
2M05350768-0536587&Orion A(N)&4&143.06&61.56&83.782024&-5.616312 \\
2M05351375-0534548&Orion A(N)&2&3.75&4.3&83.807316&-5.581916 \\
2M05351556-0509512&Orion A(N)&4&3.78&5.47&83.814867&-5.164232 \\
2M05351822-0513068&Orion A(N)&4&4.02&3.98&83.825949&-5.218559 \\
2M05352157-0509497&Orion A(N)&2&2.41&3.98&83.839905&-5.163816 \\
2M05352236-0507392&Orion A(N)&2&1.78&3.39&83.843174&-5.127558 \\
2M05353621-0520204&Orion A(N)&2&2.06&3.49&83.900904&-5.339008 \\
2M05354076-0512479&Orion A(N)&2&4.4&8.17&83.919835&-5.213314 \\
2M05354813-0553573&Orion A(N)&2&2.64&4.01&83.950564&-5.899267 \\
2M05360079-0541065&Orion A(N)&2&4.65&8.92&84.003301&-5.685141 \\
2M05362423-0544486&Orion A(N)&2&6.65&9.89&84.100972&-5.746839 \\
2M05363023-0642460&Orion A(S)&2&2.12&4.11&84.12596&-6.712795 \\
2M05364134-0634003&Orion A(S)&2&2.37&3.01&84.172283&-6.566768 \\
2M05373823-0633155&Orion A(S)&2&5.5&5.15&84.409327&-6.554327 \\
2M05384336-0701348&Orion A(S)&2&3.31&4.3&84.680707&-7.026359 \\
2M05385451-0727520&Orion A(S)&2&44.71&4.2&84.727147&-7.464459 \\
2M05392599-0804068&Orion A(S)&2&2.27&3.12&84.8583&-8.068568 \\
2M05393560-0810468&Orion A(S)&2&7.36&10.12&84.898367&-8.179668 \\
2M05393642-0700585&Orion A(S)&3&1.62&3.35&84.901791&-7.016255 \\
2M05401986-0816379&Orion A(S)&2&37.72&51.23&85.082764&-8.277201 \\
2M06402063+0940499&NGC 2264&5&1.76&3.74&100.085997&9.680554 \\
2M06403086+0934405&NGC 2264&5&5.64&3.98&100.128584&9.577939 \\
2M06404485+0957442&NGC 2264&5&29.32&53.54&100.18688&9.962293 \\
2M06405571+0951138&NGC 2264&6&28.31&33.21&100.232154&9.853848 \\
2M06405932+0946165&NGC 2264&4&4.9&6.1&100.247171&9.771273 \\
2M06411893+0927157&NGC 2264&5&8.56&4.08&100.328891&9.454382 \\
2M06413207+1001049&NGC 2264&6&75.87&40.08&100.383647&10.018033 \\
\enddata



\end{deluxetable*}

\end{document}